%% file: main.tex
\pdfoutput=1
\documentclass[sigplan,screen]{acmart}

\input{util/0-main}

\includeversion{arxiv}  
\excludeversion{asplos} 

\copyrightyear{2024}
\acmYear{2024}
\setcopyright{rightsretained}
\acmConference[ASPLOS '24]{29th ACM International Conference on Architectural Support for Programming Languages and Operating Systems, Volume 3}{April 27-May 1, 2024}{La Jolla, CA, USA}
\acmBooktitle{29th ACM International Conference on Architectural Support for Programming Languages and Operating Systems, Volume 3 (ASPLOS '24), April 27-May 1, 2024, La Jolla, CA, USA}
\acmDOI{10.1145/3620666.3651358}
\acmISBN{979-8-4007-0386-7/24/04}

\title{CSSTs: A Dynamic Data Structure for Partial Orders in Concurrent Execution Analysis}

\author[H. C. Tun\c{c}]{H\"{u}nkar Can Tun\c{c}}
\orcid{0000-0001-9125-8506} 
\affiliation{
	\institution{Aarhus University}            %% \institution is required
	\country{Denmark}                    %% \country is recommended
}
\email{tunc@cs.au.dk}          %% \email is recommended

\author[A.P. Deshmukh]{Ameya Prashant Deshmukh}
\orcid{0009-0007-6393-9188}
\affiliation{
	\institution{IIT Bombay}            %% \institution is required
	\country{India}                    %% \country is recommended
}
\email{ameyapd@cse.iitb.ac.in}          %% \email is recommended

\author[B. {\c{C}}irisci]{Berk {\c{C}}irisci}
\orcid{0000-0003-4261-090X}
\affiliation{
	\institution{Amazon Web Services}            %% \institution is required
	\country{Germany}                    %% \country is recommended
}
\email{cirisci@amazon.de}          %% \email is recommended

\author[C. Enea]{Constantin Enea}
\orcid{0000-0003-2727-8865}
\affiliation{
	\institution{Universit\'e de Paris}            %% \institution is required
	\country{France}                    %% \country is recommended
}
\email{cenea@irif.fr}          %% \email is recommended

\author[A. Pavlogiannis]{Andreas Pavlogiannis}
\orcid{0000-0002-8943-0722}
\affiliation{
	\institution{Aarhus University}            %% \institution is required
	\country{Denmark}                    %% \country is recommended
}
\email{pavlogiannis@cs.au.dk}          %% \email is recommended

\begin{document}

\input{util/after-begin-commands}

\input{abstract}

\maketitle % should come after the abstract
%\pagestyle{plain} % should come right after \maketitle

\input{sections/intro-main}
\input{sections/prelim-main}
\input{sections/ds-main}
\input{sections/ds-incremental}

\input{sections/experiments-main}
\input{sections/related_work-main}

\input{sections/conclusion}

\begin{acks}
This work was partially supported by a research grant (VIL42117) from VILLUM FONDEN.
\end{acks}

\clearpage
\appendix
\input{artifact_appendix}

%\clearpage
%\bibliographystyle{plain}
\bibliographystyle{ACM-Reference-Format}\balance
\bibliography{references}

%\clearpage
%\appendix
\begin{arxiv}
\input{appendix/0-main}
\end{arxiv}

\end{document}

%% file: util/0-main.tex
\input{util/packages}
\input{util/style}

\input{util/names}

\input{util/generic-notation}
\input{util/ds-notation}

\input{util/figures-format}
\input{util/algorithm-notation}

%% file: util/packages.tex
\usepackage{multirow, makecell}
\usepackage[utf8]{inputenc} 
\usepackage{amsmath, amsthm}
\usepackage{thmtools}
\usepackage[english]{babel}
\usepackage{pdfpages}
\usepackage{graphicx}
\usepackage{subcaption}
\usepackage{caption}
\usepackage{longtable}
\usepackage{booktabs}
\usepackage{tabu}
\usepackage[referable]{threeparttablex}
\usepackage{varwidth}
\usepackage{multirow}
\usepackage{wrapfig}
\usepackage{microtype}
\usepackage{xifthen}
\usepackage{versions}
\usepackage{dsfont}
\usepackage{xcolor}
\usepackage{pifont}
\usepackage{adjustbox}
\newcounter{index}
\usepackage{enumitem}

\usepackage[ruled,vlined,resetcount,linesnumbered]{algorithm2e}

\usepackage{hyperref}
\usepackage{thm-restate}
\usepackage[capitalise]{cleveref}

\crefname{figure}{Figure}{Figure}
\crefformat{footnote}{#2\footnotemark[#1]#3}
\usepackage{comment}

\usepackage{tikz}
\usetikzlibrary{arrows,automata,shapes,decorations,decorations.markings,calc, matrix,decorations.pathmorphing, patterns,backgrounds,shapes.misc,arrows.meta,positioning}
\usetikzlibrary{trees}
\usetikzlibrary{calc}

\usepackage{bm}
\usepackage{pgfplots}
\usepackage{enumerate}
\usepackage{pbox}

\usepackage{appendix}
\usepackage{calc}
\usepackage{fp}
\usepackage{multirow}
\usepackage{pbox}

\usepackage[mathscr]{euscript}
\usepackage{mathtools}

\usepackage{footnote}
\usepackage{bbm}
\usepackage{multicol}
\usepackage{graphicx} % omit 'demo' option in real doc.
\usepackage{soul}
\usepackage{listings, xcolor}
\usepackage{appendix}
\usepackage{versions}

\usepgfplotslibrary{groupplots}

%% file: util/style.tex
\newenvironment{myparagraph}[1]{\smallskip \noindent{\bf #1.}}{}

\newcommand{\SubParagraph}[1]{\smallskip\noindent{\em #1}}

\newtheorem{example}{Example}

\DeclareFontFamily{U}{matha}{\hyphenchar\font45}
\DeclareFontShape{U}{matha}{m}{n}{
      <5> <6> <7> <8> <9> <10> gen * matha
      <10.95> matha10 <12> <14.4> <17.28> <20.74> <24.88> matha12
      }{}
\DeclareSymbolFont{matha}{U}{matha}{m}{n}

\DeclareMathSymbol{\Lt}{3}{matha}{"CE}
\DeclareMathSymbol{\Gt}{3}{matha}{"CF}

%% file: util/names.tex
\newcommand{\mtwo}{\mathtt{M2}}
\newcommand{\seqcheck}{\mathtt{SeqCheck}}
\newcommand{\convulpoe}{\mathtt{ConVulPOE}}
\newcommand{\ufo}{\mathtt{UFO}}

\newcommand{\celeventester}{\mathtt{C11Tester}}

\newcommand{\ds}{\mathtt{CSSTs}}
\newcommand{\sparsest}{\mathtt{SST}}

\newcommand{\segtree}{\mathtt{ST}}
\newcommand{\graphds}{\mathtt{Graphs}}

\newcommand{\vc}{\mathtt{VCs}}
\newcommand{\naiveds}{\mathtt{STs}}

%% file: util/generic-notation.tex
\newcommand{\tr}{\sigma}

\newcommand{\events}[1]{\mathsf{Events}_{#1}}
\newcommand{\tho}[1]{\leq^{#1}_{\mathsf{po}}}

\newcommand{\partialord}[1]{\leq^{#1}_{\mathsf{P}}}
\newcommand{\strictord}[1]{<^{#1}_{\mathsf{P}}}

\newcommand{\set}[1]{\{#1\}}

\newcommand{\tuple}[1]{\langle #1 \rangle}

\newcommand{\Nats}{\mathbb{N}}

\newcommand{\idof}[1]{\operatorname{id}(#1)}
\newcommand{\threadof}[1]{\operatorname{tid}(#1)}

\newcommand{\rd}{\mathtt{r}}
\newcommand{\wt}{\mathtt{w}}

\newcommand{\numEvents}{n}
\newcommand{\numThreads}{k}
\newcommand{\numChainEvents}{\ell}

\newlist{compactitem}{itemize}{3} % 3 is max-depth
\setlist[compactitem]{label=\textbullet,nosep,leftmargin=*}
\crefname{compactitemi}{Item}{Items}

\newlist{compactenum}{enumerate}{3} % 3 is max-depth
\setlist[compactenum]{label=\arabic*., nosep,leftmargin=*}
\crefname{compactenumi}{Item}{Items}

\DeclareMathOperator*{\argmin}{arg\,min}

%% file: util/ds-notation.tex
\newcommand{\graphnode}[2]{{\langle #1, #2 \rangle}}
\newcommand{\directedge}{\xrightarrow{}}
\newcommand{\reach}{\mathrel{\xrightarrow{}^*}}

\newcommand{\threshold}{b}

\newcommand{\upd}{\mathtt{update}}
\newcommand{\mn}{\mathtt{min}}
\newcommand{\argleq}{\mathtt{argleq}}
\newcommand{\width}{\mathtt{width}}
\newcommand{\sub}[3]{\mathtt{Sub(#1)}_{#2}^{#3}}

\newcommand{\dsarray}[2]{\mathtt{A}_{#1}^{#2}}

\newcommand{\closure}{\mathtt{closure}}
\newcommand{\changed}{\mathtt{changed}}
\newcommand{\pred}{\mathtt{predecessor}}
\newcommand{\suc}{\mathtt{successor}}

\newcommand{\insedge}{\mathtt{insertEdge}}
\newcommand{\deledge}{\mathtt{deleteEdge}}
\newcommand{\delnode}{\mathtt{deleteNode}}
\newcommand{\del}{\mathtt{delete}}
\newcommand{\reachable}{\mathtt{reachable}}
\newcommand{\block}{\mathtt{block}}

\newcommand{\rootnode}{\mathtt{root}}

\newcommand{\start}{\mathtt{start}}
\newcommand{\en}{\mathtt{end}}
\newcommand{\leftt}{\mathtt{left}}
\newcommand{\rightt}{\mathtt{right}}

\newcommand{\node}{\mathtt{nd}}
\newcommand{\length}{\mathtt{length}}
\newcommand{\pos}{\mathtt{pos}}
\newcommand{\nil}{\mathtt{nil}}

\newcommand{\height}{\mathtt{height}}
\newcommand{\depth}{\mathtt{depth}}
\newcommand{\edgeheap}{\mathtt{edgeHeap}}

\newcommand{\insertitem}{\mathtt{insert}}

\newcommand{\MaxDegree}{\delta}
\newcommand{\CrossChainDensity}{d}
\newcommand{\AvgDensity}{q}

%% file: util/figures-format.tex
\newcommand{\mopartial}
	
\colorlet{colorPO}{darkgray!80!black}
\colorlet{colorRF}{blue}
\colorlet{colorCO}{red!80!black}
\colorlet{colorMO}{red!80!black}
\colorlet{colorFR}{purple}
\colorlet{colorECO}{orange}
\colorlet{colorCOM}{magenta}
\colorlet{colorSW}{teal}
\colorlet{colorHB}{green!40!black}
\colorlet{colorPPO}{magenta}
\colorlet{colorRSEQ}{green!40!black}
\colorlet{colorSC}{violet}
\colorlet{colorHBMO}{brown}
\colorlet{colorPSC}{violet}
\colorlet{colorREL}{olive}
\colorlet{colorWB}{olive}
\colorlet{colorRMW}{brown}

\definecolor{bred}{RGB}{213,94,0}
\definecolor{bgreen}{RGB}{0,158,115}
\definecolor{bblue}{RGB}{0,114,178}

\newcommand\numberthis{\addtocounter{equation}{1}\tag{\theequation}}
\newcommand\mathbftt[1]{\textnormal{\ttfamily\bfseries #1}}

\tikzset{
	every path/.style={>=stealth},
	po/.style={->,color=colorPO,shorten >=-0.5mm,shorten <=-0.5mm},
	ppo/.style={->,color=colorPPO,shorten >=-0.5mm,shorten <=-0.5mm},
	sw/.style={->,color=colorSW,shorten >=-0.5mm,shorten <=-0.5mm},
	rf/.style={->,color=colorRF,dotted, very thick,shorten >=-0.5mm,shorten <=-0.5mm},
	rfsolid/.style={->,color=bgreen, very thick,shorten >=-0.5mm,shorten <=-0.5mm},
	mrf/.style={->,color=colorRF,dashed,shorten >=-0.5mm,shorten <=-0.5mm},   
	urf/.style={->,color=colorRF,shorten >=-0.5mm,shorten <=-0.5mm},
	fr/.style={->,color=colorFR,dashed,shorten >=-0.5mm,shorten <=-0.5mm},
	hb/.style={->,color=bblue,thick,shorten >=-0.5mm,shorten <=-0.5mm},
	co/.style={->,color=colorCO,dotted,very thick,shorten >=-0.5mm,shorten <=-0.5mm},
	mo/.style={->,color=bred,dotted,very thick,shorten >=-0.5mm,shorten <=-0.5mm},
	mosolid/.style={->,color=bred,very thick,shorten >=-0.5mm,shorten <=-0.5mm},
	rmw/.style={->,color=colorRMW,thick,shorten >=-0.5mm,shorten <=-0.5mm},
	rseq/.style={->,color=colorRSEQ,thick,dotted,shorten >=-0.5mm,shorten <=-0.5mm},
	com/.style={->,color=colorCOM,thick,shorten >=-0.5mm,shorten <=-0.5mm},
}

\newcommand{\tikznodeopt}[4]{
	\ensuremath{
		\scalebox{0.8}{
			$\node_{#1, #2}\colon$ (#3$, #4)$ 
		}
	}
}

\newcommand{\tikznodeorig}[3]{
	\ensuremath{
		\scalebox{0.75}{
			$\node_{#1, #2}$: $#3$\hspace{-0.15cm}
		}
	}
}

\pgfplotsset{compat=1.11,
    /pgfplots/ybar legend/.style={
    /pgfplots/legend image code/.code={%
       \draw[##1,/tikz/.cd,yshift=-0.25em]
        (0cm,0cm) rectangle (3pt,0.8em);},
   },
}

%% file: util/algorithm-notation.tex
\SetKwProg{MyProcedure}{procedure}{}{}
\SetKwProg{MyFunction}{function}{}{}
\SetKwFunction{ProcNewEvent}{handleEvent}
\SetKwFunction{FunctionComputeNewOrderings}{computeOrderings}
\SetKwFunction{FunctionMaintainPO}{maintainPartialOrder}

\SetKwFunction{FunctionQuery}{{$\reachable$}}
\SetKwFunction{FunctionInsertEdge}{{$\insedge$}}
\SetKwFunction{FunctionDeleteEdge}{{$\deledge$}}
\SetKwFunction{FunctionSucc}{{$\suc$}}
\SetKwFunction{FunctionPred}{{$\pred$}}

\SetKwFunction{FunctionMin}{{$\mn$}}
\SetKwFunction{FunctionArgLeq}{{$\argleq$}}
\SetKwFunction{FunctionUpdate}{{$\upd$}}
\SetKwFunction{FunctionUpdateHelper}{{$\mathtt{updateHelper}$}}
\SetKwFunction{FunctionCreateNode}{{$\mathtt{createNode}$}}
\SetKwFunction{FunctionCreateIntermediateNodes}{{$\mathtt{createIntermediateNodes}$}}
\SetKwFunction{FunctionCreateLowestCommonAncestor}{{$\mathtt{createLowestCommonAncestor}$}}

\SetKwFunction{FunctionAddEvent}{addEvent}

\newcommand{\lIfElse}[3]{\lIf{#1}{#2 \textbf{else}~#3}}

%% file: util/after-begin-commands.tex
\newcommand*\circled[1]{\tikz[baseline=(char.base)]{
		\node[shape=circle,draw,inner sep=1pt] (char) {#1};}}

%% file: abstract.tex
\begin{abstract}
Dynamic analyses are a standard approach to analyzing and testing concurrent programs.
Such techniques observe program traces $\tr$ and analyze them to infer the presence or absence of bugs.
At its core, each analysis maintains a partial order $P$ that represents order dependencies between the events of $\tr$ .
Naturally, the scalability of the analysis largely depends on maintaining $P$ efficiently.
The standard data structure for this task has thus far been Vector Clocks.
These, however, are slow for analyses that follow a non-streaming style, costing $O(\numEvents)$ time for inserting (and propagating) each new ordering in $P$, where $\numEvents$ is the size of $\tr$, while they cannot handle the deletion of existing orderings.

In this paper we develop \emph{Collective Sparse Segment Trees} (\emph{CSSTs}), a simple but elegant data structure for maintaining a partial order $P$.
CSSTs thrive when the width $\numThreads$ of $P$ is much smaller than the size $\numEvents$ of its domain,
allowing inserting, deleting, and querying for orderings in $P$ to run in $O(\log \numEvents)$ time.
For a concurrent trace, $\numThreads$ normally equals the number of its threads, and is orders of magnitude smaller than its size $\numEvents$, making CSSTs fitting for this setting.
Our experiments confirm that CSSTs are the best data structure currently to handle a range of dynamic analyses from existing literature.
\end{abstract}


\ccsdesc[500]{Software and its engineering~Software verification and validation}
\ccsdesc[300]{Theory of computation~Theory and algorithms for application domains}
\ccsdesc[300]{Theory of computation~Program analysis}

\keywords{concurrency, happens-before, vector clocks, dynamic concurrency analyses, dynamic reachability}

%\begin{abstract}
%In this paper we devise a data structure for efficiently maintaining partial orders.
%Partial orders are a standard mathematical tool in the context of analyzing concurrent programs. 
%In this setting, a recurring sub-problem is to devise efficient methods to compute reachability in a
%given partial order [13, 49, 56]. Previous works have often utilized ad hod structures to address this
%problem. Developing a standard method which demolishes the need for such ad hod structures is
%an open problem. In this work, we target the problem of solving the reachability problem in partial
%orders with small width. We formulate this problem as a dynamic reachability problem on DAGs. The
%task is to answer each reachability query correctly, accounting for all preceding edge/node insertions
%and deletions. We have identified several problem domains in which our approach can be applied,
%and we have substituted existing techniques with our novel method
%\end{abstract}

%% file: sections/intro-main.tex
%!TEX root = main.tex

%\input{\intropath/1-beginning}
%\input{\intropath/2-motivating}
%\input{\intropath/3-contributions}

\input{sections/intro-beginning}
\input{sections/intro-motivating}
\input{sections/intro-contributions}

%% file: sections/intro-beginning.tex
\section{Introduction}\label{sec:intro}

Concurrent programming is notoriously difficult due to the inherent nondeterminism in interprocess communication~\cite{Musuvathi2008}.
As such, considerable efforts go towards effective techniques for analyzing concurrent programs and testing them for correctness.
Dynamic analyses are one of the most widespread types of such techniques.
Their main working principle  is to discover software faults by analyzing concrete program executions.
%and reasoning whether a bug has taken place, or even could have taken place under some alternative but valid execution.
Prominent examples include concurrency bugs such as data races~\cite{Flanagan-2009,Eizenberg-2017,Roemer-2018,Mathur-2018,Pavlogiannis-2019,Mathur-2020b,Roemer-2020,Luo-2021,Gao-2023}, deadlocks~\cite{Kalhauge-2018,Cai-2021,Tunc-2023}, linearizability~\cite{Cirisci-2020} and atomicity violations~\cite{Flanagan-2008,Biswas-2014,Mathur-2018,Cai-2021}, thread-safety~\cite{Li-2019}, and memory violations~\cite{Huang-2018,Yu-2021}, to name a few.

Although each analysis employs reasoning that is specific to the problem at hand, one of its core components is nearly always the construction of a partial order $P$ (often termed generically as ``happens-before''~\cite{Lamport-1978}) that captures order dependencies between events of the analyzed trace $\tr$.
The analysis undergoes a sequence of interleaved operations of 
(i)~\emph{inserting} new orderings $e_1\to e_2$ in $P$, 
(ii)~\emph{querying} whether $e_1\to e_2$ in $P$, and even
(iii)~\emph{deleting} existing orderings $e_1\to e_2$ from $P$,
as it attempts to prove that $\tr$ has certain properties (e.g., that it is sequentially-consistent, or it contains a data race).
To benefit scalability, each operation must take as little time as possible, as both the number of traces and the size of each trace normally span several orders of magnitude.

\input{material/examples/motivating}

Consider analyzing a trace $\tr$ of size $\numEvents$.
Representing $P$ as a graph $G$, the above setting is colloquially known as \emph{dynamic graph reachability}, with dynamically inserting/deleting edges and performing reachability queries.
Such queries are handled in  $O(1)$ time, at the cost of $O(\numEvents^2)$ for keeping $G$ transitively closed after each update.
This bound is prohibitively large for typical values of $\numEvents$ in the range $10^{4}$-$10^{10}$.

The scalability of dynamic analyses has been driven primarily by a simple data structure known as \emph{Vector Clocks}~\cite{Mattern-1989}, based on the realization that
(i)~$\tr$ consists of a small number of threads $\numThreads$, and
(ii)~normally, $P$ contains $\numThreads$ totally ordered chains (or a number proportional to $\numThreads$).
Consequently, the whole backward set of an event $e$ can be compactly summarized as an integer array, i.e., a Vector Clock.
This allows to replace one factor $\numEvents$ in the update complexity by the much smaller $\numThreads$, i.e., a single edge insertion now takes $O(\numEvents \numThreads)$ time.

In many analyses that have a \emph{streaming nature}, the insertion cost of Vector Clocks is further reduced to $O(\numThreads)$.
Here the events of $\tr$ are processed one by one, and new orderings $e_1\to e_2$ are always inserted where $e_2$ is the current event being processed. 
Popular examples include analyses for detecting data-races~\cite{Flanagan-2009,Mathur-2018,Roemer-2018}, deadlocks~\cite{Tunc-2023}, and atomicity violations~\cite{Mathur-2020}, as well as consistency checking in some weak-memory models~\cite{Tunc-2023b}.
Thus in the streaming setting, Vector Clocks are arguably the best data structure to represent a partial order.
On the other hand, many analyses are \emph{non-streaming}, i.e., as the analysis progresses, it might insert new orderings between arbitrary events of $\tr$.
These new orderings might have to be propagated across $n$ events, recovering the $O(\numEvents \numThreads)$ bound, which can lead to a significant slowdown.
A natural question thus arises:~\emph{can the $O(\numEvents \numThreads)$ cost be reduced further?}
In general, \emph{what is the best data structure for non-streaming concurrent execution analysis?}

Here we develop a simple but elegant data structure for this general setting, called \emph{Collective Sparse Segment Trees} ($\ds$).
Like Vector Clocks, $\ds$ exploit the fact that $\numThreads\Lt \numEvents$, and offer update and querying operations that run roughly in $O(\log \numEvents)$ time, thereby reducing the prohibitive linear dependency on $\numEvents$ to logarithmic.
We argue that, as Vector Clocks are the most fitting data structure for handling \emph{streaming} dynamic analyses, $\ds$ are the most fitting data structure for handling non-streaming analyses.

%% file: material/examples/motivating.tex
%!TEX root = ../../main.tex

\begin{figure*}[!h]
	\centering
	\def\ystep{0.5}
	\begin{subfigure}[t]{0.31\textwidth}
		\scalebox{0.95}{
			\begin{tikzpicture}[yscale=1]
				\node (t10) at (0,1.2*\ystep) {Thread 0};
				\node (t11) at (0,0*\ystep) {$e_0\colon \wt(x, 1)$};
				\node (t12) at (0,-2*\ystep) {$e_1\colon \rd(y, 5)$};
				\node (t13) at (0,-4*\ystep) {\colorbox{gray!20}{{$e_2: \rd(x,3)$}} };
				
				\node (t20) at (2,1.2*\ystep) {Thread 1};
				\node (t21) at (2,0*\ystep) {$e_3\colon \wt(x, 3)$};
				\node (t22) at (2,-2*\ystep) {$e_4\colon \wt(y, 4)$};
				\node (t23) at (2,-4*\ystep) {$e_5\colon \wt(y, 5)$};
				
				\node (t30) at (4.0,1.2*\ystep) {Thread 2};
				\node (t31) at (4.0,0*\ystep) {$e_6\colon \wt(x, 3)$};
				\node (t32) at (4.0,-2*\ystep) {$\ldots$};
				\node (t33) at (4.0,-4*\ystep) {$e_n\colon \rd(y, 4)$};
				
				%mo edges
				\draw[po] (t21) to (t31);
				
				%rf edges
				\draw[hb, bend right=15, pos=0.1] (t23) to (t12);
				\draw[hb] (t22) to (t33);
				
				%fr edges
				\draw[po] (t33) to (t23);
				
				%po edges
				\draw[po] (t11) to (t12);
				\draw[po] (t21) to (t22);
				\draw[po] (t22) to (t23);
				\draw[po] (t31) to (t32);
				\draw[po] (t32) to (t33);
				
				%new edges
				\draw[rfsolid,bend left=0] (t12) to node[left, pos=0.5]{$\circled{1}$} (t13);
			\end{tikzpicture} 
		}
		\caption{
			Initial partial order.
		}
		\label{fig:motivating-1}
	\end{subfigure}
	\hfill
	\begin{subfigure}[t]{0.31\textwidth}
		\scalebox{0.95}{
			\begin{tikzpicture}[yscale=1]
				\node (t10) at (0,1.2*\ystep) {Thread 0};
				\node (t11) at (0,0*\ystep) {$e_0\colon \wt(x, 1)$};
				\node (t12) at (0,-2*\ystep) {$e_1\colon \rd(y, 5)$};
				\node (t13) at (0,-4*\ystep) {\colorbox{gray!20}{{$e_2\colon \rd(x,3)$}} };
				
				\node (t20) at (2,1.2*\ystep) {Thread 1};
				\node (t21) at (2,0*\ystep) {$e_3\colon \wt(x, 3)$};
				\node (t22) at (2,-2*\ystep) {$e_4\colon \wt(y, 4)$};
				\node (t23) at (2,-4*\ystep) {$e_5\colon \wt(y, 5)$};
				
				\node (t30) at (4.0,1.2*\ystep) {Thread 2};
				\node (t31) at (4.0,0*\ystep) {$e_6\colon \wt(x, 3)$};
				\node (t32) at (4.0,-2*\ystep) {$\ldots$};
				\node (t33) at (4.0,-4*\ystep) {$e_n\colon \rd(y, 4)$};
				
				%mo edges
				\draw[po] (t21) to (t31);
				
				%rf edges
				\draw[hb, bend right=15, pos=0.1] (t23) to (t12);
				\draw[hb] (t22) to (t33);
				
				%fr edges
				\draw[po] (t33) to (t23);
				
				%po edges
				\draw[po] (t11) to (t12);
				\draw[po] (t12) to (t13);
				\draw[po] (t21) to (t22);
				\draw[po] (t22) to (t23);
				\draw[po] (t31) to (t32);
				\draw[po] (t32) to (t33);
				
				%new edges
				\draw[hb,bend right=5] (t21) to node[left, pos=0.1]{$\circled{2}$~~} (t13);
				
				\draw[mosolid] (t11) to node[above]{$\circled{3}$} (t21);
				\draw[mosolid, bend right=16] (t13) to node[left, pos=0.9]{$\circled{4}$~~}  (t31);
			\end{tikzpicture}  
		}
		\caption{
			First update.
		}
		\label{fig:motivating-2}
	\end{subfigure}
	\hfill
	\begin{subfigure}[t]{0.31\textwidth}
		\scalebox{0.95}{
			\begin{tikzpicture}[yscale=1]
				\node (t10) at (0,1.2*\ystep) {Thread 0};
				\node (t11) at (0,0*\ystep) {$e_0\colon \wt(x, 1)$};
				\node (t12) at (0,-2*\ystep) {$e_1\colon \rd(y, 5)$};
				\node (t13) at (0,-4*\ystep) {\colorbox{gray!20}{{$e_2\colon \rd(x,3)$}} };
				
				\node (t20) at (2,1.2*\ystep) {Thread 1};
				\node (t21) at (2,0*\ystep) {$e_3\colon \wt(x, 3)$};
				\node (t22) at (2,-2*\ystep) {$e_4\colon \wt(y, 4)$};
				\node (t23) at (2,-4*\ystep) {$e_5\colon \wt(y, 5)$};
				
				\node (t30) at (4.0,1.2*\ystep) {Thread 2};
				\node (t31) at (4.0,0*\ystep) {$e_6\colon \wt(x, 3)$};
				\node (t32) at (4.0,-2*\ystep) {$\ldots$};
				\node (t33) at (4.0,-4*\ystep) {$e_n\colon \rd(y, 4)$};
				
				%mo edges
				\draw[po] (t21) to (t31);
				
				%rf edges
				\draw[hb, bend right=15, pos=0.1] (t23) to (t12);
				\draw[hb] (t22) to (t33);
				
				%fr edges
				\draw[po] (t33) to (t23);
				
				%po edges
				\draw[po] (t11) to (t12);
				\draw[po] (t12) to (t13);
				\draw[po] (t21) to (t22);
				\draw[po] (t22) to (t23);
				\draw[po] (t31) to (t32);
				\draw[po] (t32) to (t33);
				
				%new edges
				\draw[hb, bend left=16] (t31) to node[above, pos=0.25]{$\circled{5}$~~} (t13);
				
				\draw[mosolid, bend left=12] (t11) to node[above, pos=0.05]{$\circled{6}$} (t31);
				
			\end{tikzpicture} 
		}
		\caption{
			Delete and update.
		}
		\label{fig:motivating-3}
	\end{subfigure}
	\caption{
		A consistency analysis. Blue edges represent reads-from information. Numbered edges represent new orderings inserted by the analysis.
	}
	\label{fig:motivating}
\end{figure*}

%% file: sections/intro-motivating.tex
\subsection{Motivating Example} \label{sec:motivating}

Consider the analysis of an abstract trace shown in \cref{fig:motivating}, consisting of three threads and write/read events $\wt(x,v)$/$\rd(x,v)$, where $x$ is a variable and $v$ is a value.
The consistency-testing problem asks whether there is an interleaving that is consistent with the values observed by each read~\cite{Gibbons-1997}, and has numerous applications in dynamic concurrency analyses~\cite{Chalupa-2018,Bui-2021,Luo-2021,Roy-2006,Emmi-2019,Tunc-2023,Agarwal-2021,Huang-2018,Kalhauge-2018,Chakraborty-2024a}.
The analysis maintains 
(i)~a reads-from map $\wt(x,v)\mapsto \rd(x,v)$, denoting that $\rd(x,v)$ obtains its value from $\wt(x,v)$, and
(ii)~a partial order $P$ that is induced by this reads-from map and any additional indirect orderings this might impose.

Assume that the consistency analysis has, so far, established the partial order $P$ in \cref{fig:motivating-1}, with blue edges marking the current reads-from map.
The analysis proceeds on the read event $e_2\colon \rd(x,3)$. 
It first inserts in $P$ the program order edge $\color{bgreen}{\circled{1}}$. 
As $e_2$ has no successors in $P$, this edge need not be propagated further.
Now the analysis must decide which of the two writes $e_3$ and $e_6$ can be observed by $e_2$.
In general, there is no efficient procedure for deciding this~\cite{Gibbons-1997}, and the consistency algorithm proceeds by trying each option.

In \cref{fig:motivating-2}, the analysis tries to map $e_3\mapsto e_2$, leading to inserting $\color{bblue}{\circled{2}}$ in $P$.
Moreover, since $e_0\to e_2$ in $P$, the analysis also inserts $\color{bred}{\circled{3}}$ and $\color{bred}{\circled{4}}$ in $P$.
These are necessary orderings to respect the fact that $e_2$ reads from $e_3$ --- the process of inferring such orderings is known as saturation, and is used widely in dynamic analyses (e.g.,~\cite{Roy-2006,Pavlogiannis-2019,Roemer-2018,Abdulla-2019,Zennou-2019,Zennou-2020,Luo-2021}).
Crucially, both $e_3$ and $e_6$ have many successors in $P$, thus these edges must be propagated along many events, costing $O(\numEvents)$ time, even if $P$ is represented using Vector Clocks.
In contrast, $\ds$ achieve a much smaller $O(\log n)$ bound.

The reads-from choice $e_3\mapsto e_2$ is inconsistent due to the cycle $e_2\to e_6\to\dots\to e_n\to e_5\to e_1 \to e_2$.
The analysis must delete the orderings $\color{bblue}{\circled{2}}$, $\color{bred}{\circled{3}}$ and $\color{bred}{\circled{4}}$, before proceeding with an alternative writer for $e_2$.
When $P$ is represented via Vector Clocks, there is no efficient method for deletion, and essentially it must be recomputed from scratch, taking $O(\numEvents^2)$ time.
In contrast, $\ds$ only need $O(\log \numEvents)$ time per edge deletion, thereby completing this task efficiently.

Finally, in \cref{fig:motivating-3}, the analysis tries to map $e_6\mapsto e_2$, leading to inserting $\color{bblue}{\circled{5}}$ in $P$.
It also infers ordering $\color{bred}{\circled{6}}$, which again must be propagated further, a task that costs $O(\numEvents)$ time for Vector Clocks but only $O(\log \numEvents)$ time for $\ds$.

%% file: sections/intro-contributions.tex
\subsection{Contributions}\label{subsec:contributions}

We develop $\ds$, a new data structure for maintaining and querying partial orders of small width.
Concretely, consider the maintenance of a DAG $G$, representing a partial order $P$ over $\numEvents$ events and $\numThreads$ totally ordered chains (normally $\numThreads$ equals the number of threads).
\begin{compactenum}
\item $\ds$ support fully dynamic updates (inserting and deleting edges) in $O(\max(\log \MaxDegree,\min(\log \numEvents,\CrossChainDensity)))$ time, where $\MaxDegree$ is the maximum degree in $G$
and $\CrossChainDensity$ is the \emph{cross-chain density} of $G$, a novel notion we introduce here that captures a type of sparsity prevalent in practice.
Normally, $\MaxDegree$ is constant and the above expression results in $O(\CrossChainDensity)$.
Reachability (ordering) queries take $O(\numThreads^3\min(\log n, \CrossChainDensity))$ time.
\item We further optimize $\ds$ for the purely incremental setting, supporting the insertion, but not deletion, of edges.
Incremental $\ds$ run updates and queries in time $O(\numThreads^2 \min(\log \numEvents, d))$ and $O(\min(\log \numEvents, d))$, respectively,
\item We make an extensive experimental evaluation of $\ds$ in concurrency analyses from the literature, spanning
data-race detection~\cite{Pavlogiannis-2019},
deadlock detection~\cite{Cai-2021},
memory bugs~~\cite{Yu-2021},
consistency checking for X86-TSO~\cite{Roy-2006},
use-after-free bugs~\cite{Huang-2018},
data-races in the C11 memory model~\cite{Luo-2021},
and root cause analysis of linearizability violations~\cite{Cirisci-2020}.
Our results show that $\ds$ speed up the corresponding analysis in most cases,
indicating that they are the most fitting data structure in this ubiquitous setting.
\end{compactenum}

$\ds$ are based on the insight that dynamic reachability on partial orders of small width can be reduced to a basic algorithmic problem known as \emph{dynamic suffix minima}.
This idea was partly explored in a data structure for incremental reachability underpinning the M2 data-race detector~\cite{Pavlogiannis-2019}.
However, $\ds$ are more advanced, as
(i)~they employ techniques making them faster in all cases,
(ii)~they achieve tighter time and space bounds on sparse instances, and
(iii)~they support fully dynamic updates (i.e., both incremental and decremental).
Our experimental results confirm that $\ds$ are indeed more efficient in both time and memory, besides being broader (as they handle edge deletions).
\begin{asplos}
In our technical report~\cite{arxiv}, we provide the proofs for the theorems and lemmas presented in the paper.
\end{asplos}

%% file: sections/prelim-main.tex
%!TEX root = ../../main.tex

\section{Preliminaries}

\input{sections/prelim-basic}

%% file: sections/prelim-basic.tex
In this section, we develop some general notation and introduce standard
concepts on concurrent executions and partial orders that will be used throughout the paper.

\subsection{Concurrent Execution Model}

\myparagraph{Events and traces}
We broadly consider traces $\tr$ of concurrent programs, representing the history of operations in an execution.
We write $\events{\tr}$ to denote the set of events occurring in $\tr$.
Each event of $\tr$ is a tuple $e = \tuple{t, i, m}$, where $t$ is an identifier of the thread that performed $e$ and $i$ is the sequence id of $e$. 
The pair $\tuple{t, i}$ serves as a unique identifier for $e$.
The field $m$ represents meta information for $e$ that might be relevant to a dynamic analysis, but is immaterial for $\ds$.
E.g., $m$ may record the operation performed by $e$ (read, write, etc), as well as the variable it accesses and the value it acquires.
$\ds$ operate only on the identifier $\tuple{t, i}$, while the meta information $m$ will be used only in our examples.
We write $\threadof{e}$, $\idof{e}$ for the thread identifier and sequence id of $e$, respectively.
Depending on the application, the events in $\tr$ may or may not be totally ordered.

\myparagraph{Relations and functions on execution traces}
For a binary relation $X$, we denote by $X^*$ its reflexive transitive closure. 
A partial order is a reflexive, transitive, and anti-symmetric binary relation defined on the elements of a given set $S$.
We denote by $\partialord{\tr}$ a partial order $\mathsf{P}$ over the set $\events{\tr}$.
Given $e_i, e_j \in \events{\tr}$, we write $e_i \partialord{\tr} e_j$ to represent $(e_i, e_j) \in \partialord{\tr}$.
We write $e_i \strictord{\tr} e_j$ to represent $e_i \partialord{\tr} e_j$ and $e_i \neq e_j$.
%If both $e_i \strictord{\tr} e_j$ and $e_j \strictord{\tr} e_i$ do not hold, then we say that $(e_i, e_j)$ are unordered in $\partialord{\tr}$, denoted by $e_i \unord{\tr}{P} e_j$.
The \emph{program order} $\tho{\tr}$ represents a partial order where $e_i \tho{\tr} e_j$ iff $\idof{e_i} \leq \idof{e_j}$ and $\threadof{e_i} = \threadof{e_j}$.
We write $\width(P)$ to represent the largest size of a set $S \subseteq \events{\tr}$ such that for all distinct pairs $e_i, e_j \in S$, we have that $e_i \not\strictord{\tr} e_j$ and $e_j \not\strictord{\tr} e_i$.

In concurrency analyses, $\width(P)$ is almost always bounded by some constant, a fact that is exploited by Vector Clocks, as well as the CSSTs we introduce in this work.
E.g., in many analyses $\tho{\tr}\subseteq \partialord{\tr}$, implying that $\width(P)\leq \numThreads$, where $\numThreads$ is the number of threads.
Although in other settings (e.g., those involving weak-memory concurrency) $\width(P)$ might be larger than $\numThreads$, it still remains bounded by some small factor of $\numThreads$.
For example, analyses involving TSO normally satisfy $\width(P)\leq 2\numThreads$ (we touch on this later in \cref{sec:experiments}).

\subsection{Dynamic Reachability on DAGs}

\myparagraph{Chain directed acyclic graphs} 
We consider directed acyclic graphs (DAGs) $G = (V, E)$, where $V$ is the set of nodes and $E$ is the set of edges.
Dynamic analyses use such DAGs to represent partial orders on $\numEvents$ events of a concurrent execution, where $\numThreads$ is normally the number of threads, or a value proportional to that.
For this purpose, a node in $V$ is a pair $u=\tuple{t,i}\in [\numThreads]\times [\numEvents]$, where $[j]=\set{0,1,\ldots,j-1}$.
The set of edges normally captures the program order, which is reflected in the fact that for any two nodes $\graphnode{t}{i}, \graphnode{t}{i+1}\in V$,
we have $(\graphnode{t}{i}, \graphnode{t}{i+1})\in E$.
We can thus represent $G$ as $\numThreads$ totally-ordered chains with additional edges across chains,
and call such graphs \emph{chain DAGs}.
A node $\graphnode{t}{i}$ is said to be \emph{earlier} than another node in the same chain $\graphnode{t}{j}$ if $i \leq j$, and \emph{later} otherwise. 
Then, $G$ can be written as the composition of $\numThreads(\numThreads-1)$ subgraphs $\sub{G}{t_1}{t_2}$  induced by the chains $t_1$ and $t_2$, as well as the cross-chain edges from $t_1$ to $t_2$
(see~\cref{fig:defs-subgraph}).
The \emph{cross-chain density} $\CrossChainDensity$ of $G$ is the maximum, among all chains $t_1$, number of nodes in chain $t_1$ that have an outgoing edge to some other chain.
\input{material/examples/subgraph_def}

\myparagraph{Dynamic reachability}
The dynamic reachability problem is defined with respect to a given chain DAG $G = (V , E)$ and online sequence of operations of the following types.
\begin{compactenum}
	\item An $\insedge(u, v)$ operation, such that $(u, v) \not \in E$, inserts the edge $(u, v)$ in $G$.
	\item An $\deledge(u, v)$ operation, such that $(u, v) \in E$, deletes the edge $(u, v)$ in $G$.
	\item A $\reachable(u, v)$ operation returns True iff $u \reach v$.
	\item A $\suc(u, t)$ operation returns the earliest successor of $u$ in the $t$-th chain.
	\item A $\pred(u, t)$ operation returns the latest predecessor of $u$ in the $t$-th chain.
\end{compactenum}
The operations $\insedge$ and $\deledge$ are updates, while $\reachable$, $\suc$ and $\pred$ are queries. 
We allow update operations only across nodes in different chains.
In this dynamic setting, we define the cross-chain density of $G$ to be the maximum cross-chain density of the graph across all update operations on $G$.

%% file: material/examples/subgraph_def.tex
%!TEX root = ../../main.tex

\begin{figure}
\centering
\def\ystep{0.8}
\begin{subfigure}[t]{0.15\textwidth}
\centering
\scalebox{0.95}{
\definecolor{offwhite}{HTML}{000000}
\tikzset{
> = stealth,
every node/.append style = {
text = offwhite
},
every path/.append style = {
arrows = ->,
draw = offwhite,
fill = offwhite
},
hidden/.style = {
draw = offwhite,
shape = circle,
inner sep = 1pt
}
}
\tikz{
%\node (a) at (0,-0.3) {Chain $0$};
\node (a0) at (0,-1*\ystep) {$\graphnode{0}{0}$};
\node (a1) at (0,-2*\ystep) {$\graphnode{0}{1}$};
\node (a2) at (0,-3*\ystep) {$\graphnode{0}{2}$};
\path (a0) edge (a1);
\path (a1) edge (a2);

%\node (b) at (2,-0.3) {Chain $1$};
\node (b0) at (1.5,-1*\ystep) {$\graphnode{1}{0}$};
\node (b1) at (1.5,-2*\ystep) {$\graphnode{1}{1}$};
\node (b2) at (1.5,-3*\ystep) {$\graphnode{1}{2}$};
\path (b0) edge (b1);
\path (b1) edge (b2);
}
}
\caption{$\sub{G}{0}{1}$.}
\label{subfig:defs-subgraph-1}
\end{subfigure}
\hfill
\begin{subfigure}[t]{0.15\textwidth}
\centering
\scalebox{0.95}{
\definecolor{offwhite}{HTML}{000000}
\tikzset{
> = stealth,
every node/.append style = {
text = offwhite
},
every path/.append style = {
arrows = ->,
draw = offwhite,
fill = offwhite
},
hidden/.style = {
draw = offwhite,
shape = circle,
inner sep = 1pt
}
}
\tikz{
%\node (a) at (0,-0.3) {Chain $0$};
\node (a0) at (0,-1*\ystep) {$\graphnode{0}{0}$};
\node (a1) at (0,-2*\ystep) {$\graphnode{0}{1}$};
\node (a2) at (0,-3*\ystep) {$\graphnode{0}{2}$};
\path (a0) edge (a1);
\path (a1) edge (a2);

%\node (b) at (2,-0.3) {Chain $1$};
\node (b0) at (1.5,-1*\ystep) {$\graphnode{1}{0}$};
\node (b1) at (1.5,-2*\ystep) {$\graphnode{1}{1}$};
\node (b2) at (1.5,-3*\ystep) {$\graphnode{1}{2}$};
\path (b0) edge (b1);
\path (b1) edge (b2);

\path (b2) edge (a1);
}
}
\caption{$\sub{G}{1}{0}$.}
\label{subfig:defs-subgraph-2}
\end{subfigure}
\hfill
\begin{subfigure}[t]{0.15\textwidth}
\centering
\scalebox{0.95}{
\definecolor{offwhite}{HTML}{000000}
\tikzset{
> = stealth,
every node/.append style = {
text = offwhite
},
every path/.append style = {
arrows = ->,
draw = offwhite,
fill = offwhite
},
hidden/.style = {
draw = offwhite,
shape = circle,
inner sep = 1pt
}
}
\tikz{
%\node (a) at (0,-0.3) {Chain $0$};
\node (a0) at (0,-1*\ystep) {$\graphnode{1}{0}$};
\node (a1) at (0,-2*\ystep) {$\graphnode{1}{1}$};
\node (a2) at (0,-3*\ystep) {$\graphnode{1}{2}$};
\path (a0) edge (a1);
\path (a1) edge (a2);

%\node (b) at (2,-0.3) {Chain $1$};
\node (b0) at (1.5,-1*\ystep) {$\graphnode{2}{0}$};
\node (b1) at (1.5,-2*\ystep) {\ldots};
\node (b2) at (1.5,-3*\ystep) {$\graphnode{2}{n-6}$};
\path (b0) edge (b1);
\path (b1) edge (b2);

\path (a0) edge (b0);
\path (a1) edge (b2);
}
}
\caption{$\sub{G}{1}{2}$.}
\label{subfig:defs-subgraph-3}
\end{subfigure}
\caption{
Three subgraphs of the chain DAG $G$ formed from the execution in \cref{fig:motivating-1}.
The cross-chain density of $G$ is $2$, as chain $1$ (i.e., thread~1 in \cref{fig:motivating-1}) has two outgoing edges to chain $2$ (from events $e_3$ and $e_4$).
}
\label{fig:defs-subgraph}
\end{figure}

%% file: sections/ds-main.tex
\input{sections/ds-beginning}

\input{sections/ds-dynamic-suffix-minima}
\input{sections/ds-sparse-dense-segment-tree}
\input{sections/ds-cssts}

%% file: sections/ds-beginning.tex
\section{Collective Sparse Segment Trees}

We now present  \emph{Collective Sparse Segment Trees} ($\ds$), a data structure for efficiently solving the dynamic reachability problem in the context of concurrent execution analysis.
The properties of $\ds$ are stated in the following theorem.

\begin{restatable}{theorem}{thmdsdynamic}\label{thm:ds_dynamic}
Consider a chain DAG $G$ of $\numEvents$ nodes, $\numThreads$ chains, maximum out-degree $\MaxDegree$ and cross-chain density $\CrossChainDensity$.
$\ds$ maintain $G$ under dynamic updates, costing $O(\max(\log \MaxDegree,\min(\log \numEvents,\CrossChainDensity)))$ time per update and $O(\numThreads^3 \min(\log \numEvents, \CrossChainDensity))$ time per query.
\end{restatable}

$\ds$ are based on the insight that dynamic reachability on DAGs with few chains can be efficiently reduced to repeated instances of another basic problem known as \emph{dynamic suffix minima}. 
We begin by developing this concept when $G$ consists of only $\numThreads=2$ chains in \cref{subsec:dynamic-suffix-minima}.
Dynamic suffix minima are solved using a data structure known as \emph{Segment Trees}.
A key novelty of $\ds$ is based on the realization that most direct orderings in dynamic analyses make $G$ have low cross-chain density (i.e., the cross-chain edges are sparse).
We exploit this insight in \cref{subsec:sparse_segment_tree}, by developing \emph{Sparse Segment Trees} that handle arbitrary graphs, but become more efficient the lower their cross-chain density becomes.
Finally, in \cref{subsec:ds-cssts} we handle DAGs of arbitrarily many chains $\numThreads$, by appropriately maintaining \emph{collections} of Sparse Segment Trees,
which also explains the name of the data structure.

%% file: sections/ds-dynamic-suffix-minima.tex
\subsection{Dynamic Suffix Minima and Segment Trees}\label{subsec:dynamic-suffix-minima}

Here we state the algorithmic problem of dynamic suffix minima, and relate it to dynamic reachability on chain DAGs, focusing on the special case of $\numThreads=2$ chains.
In later sections we explain how to handle $\numThreads>2$.
Some insights are based on observations made in~\cite{Pavlogiannis-2019} for the purely incremental case.

\myparagraph{Dynamic suffix minima}
The dynamic suffix minima problem concerns maintaining an array $\dsarray{}{}$ of $\numEvents$ values in $\Nats\cup \{\infty\}$ under an online sequence of updates, and
answering minima queries on ranges of $\dsarray{}{}$.
We write $\dsarray{}{}[i]$ and $\dsarray{}{}[i : j]$ to denote the value of $\dsarray{}{}$ at index $i$ and the sequence of values $\dsarray{}{}[i],\ldots, \dsarray{}{}[j]$, respectively,
and write $\dsarray{}{}[i : ]$ as shorthand for $\dsarray{}{}[i : |\dsarray{}{}|-1]$.
We say that $\dsarray{}{}$ is empty on index $i$ if $\dsarray{}{}[i] = \infty$.
The density of $\dsarray{}{}$ is the number of non empty entries of $\dsarray{}{}$.

For $i\in[\numEvents]$ and $a \in \Nats\cup \{\infty\}$, an operation on $\dsarray{}{}$ is one of:
\begin{compactenum}
\item $\mn(\dsarray{}{}, i)$, returning $\mn_{i \leq j < n} \dsarray{}{}[j]$, i.e., the minimum value in the suffix $\dsarray{}{}[i:]$;
\item $\argleq(\dsarray{}{}, a)$, returning $\max_{i \colon \dsarray{}{}[i] \leq a} i$, i.e., the largest index $i$ of $\dsarray{}{}$ such that $\dsarray{}{}[i] \leq a$; and
\item $\upd(\dsarray{}{}, i, a)$,  setting $\dsarray{}{}[i] = a$.
\end{compactenum}
The task is to answer $\mn$ and $\argleq$ queries, taking into consideration all preceding $\upd$ operations.

Dynamic suffix minima (in fact, the more general problem of dynamic \emph{range} minima) is solved efficiently using Segment Trees ($\segtree$s), taking $O(\log \numEvents)$ time per operation (see, e.g.,~\cite{Arge-2013}).
A $\segtree$ $T$ represents $A$ as a binary tree of nodes $\node_{i, j}$ each associated with  a range $[i, j]$.
Each $\node_{i, j}$ maintains a value $\node_{i, j}.\mn=\min(\dsarray{}{}[i : j])$.
We write $\depth(\node_{i, j})$ to denote the depth of $\node_{i, j}$ in $T$.
We let $T.\rootnode$ be the root node of $T$.
Each query is answered by traversing $T$ from the root to a leaf, the latter containing the value to be returned.

\input{material/examples/segtree}

\myparagraph{Dynamic reachability on $\numThreads=2$ chains}
Recall that a chain DAG $G$ of $\numThreads$ chains is the composition of $\numThreads(\numThreads-1)$ subgraphs $\sub{G}{t_1}{t_2}$ (\cref{fig:defs-subgraph}).
Now assume that $\numThreads=2$.
We represent each $\sub{G}{t_1}{t_2}$ as an array $\dsarray{t_1}{t_2}\colon [n] \xrightarrow{} [n] \cup {\infty}$.
$\dsarray{t_1}{t_2}[j_1]$ stores a unique neighbor of node $\graphnode{t_1}{j_1}$ in chain $t_2$.
What if $\graphnode{t_1}{j_1}$ has multiple neighbors to $t_2$?
Then, it suffices to store the \emph{earliest neighbor} of $\graphnode{t_1}{j_1}$ in $t_2$.
Although this is not an exact representation of $G$, it captures exactly all reachability in $G$.
In particular, we maintain the invariant
\[
\dsarray{t_1}{t_2} = \min(\{j_2\colon \graphnode{t_1}{j_1} \directedge \graphnode{t_2}{j_2}\}) \numberthis \label{eq:invariant1}
\]
under the standard convention that $\min(\emptyset)=\infty$.
A query $\suc(\graphnode{t_1}{j_1}, t_2)$ simply returns the suffix minima $x=\mn(\dsarray{t_1}{t_2}, j_2)$, while  a query $\pred(\graphnode{t_2}{j_2}, t_1)$ returns the suffix arg-minima $\argleq(\dsarray{t_1}{t_2}, j_2)$.
For reachability queries, observe that we have $\graphnode{t_1}{j_1} \reach \graphnode{t_2}{j_2}$ in $G$  iff there is an edge $\graphnode{t_1}{i_1}\to \graphnode{t_2}{i_2}$ such that $i_1\geq j_1$ and $i_2\leq j_2$.
Hence, a query $\reachable(\graphnode{t_1}{j_1}, \graphnode{t_2}{j_2})$ reduces to checking whether $\suc(\graphnode{t_1}{j_1}, t_2)\leq j_2$.

How can we maintain the invariant \cref{eq:invariant1} under edge insertions and deletions?
We achieve this by storing all current successors of $\graphnode{t_1}{j_1}$ to chain $t_2$ in a min heap $\edgeheap_{t_1}^{t_2}[j_1]$.
Inserting and deleting edges from $\graphnode{t_1}{j_1}$ is handled by applying  the same operation in $\edgeheap_{t_1}^{t_2}[j_1]$.
This ensures that the earliest neighbor of $\graphnode{t_1}{j_1}$ in $t_2$ appears in the root of $\edgeheap_{t_1}^{t_2}[j_1]$, which is also stored in $\dsarray{t_1}{t_2}[j_1]$.

\input{material/examples/subgraph}

\myparagraph{Handing $\numThreads>2$ chains}
%At this point, a natural question is how to handle $\numThreads>2$ chains.
Naturally, when $\numThreads>2$, we can have a suffix minima array $\dsarray{t_1}{t_2}$ for every pair of threads $t_1,t_2\in[\numThreads]$ with $t_1\neq t_2$.
However, a path $\graphnode{t_1}{j_1} \reach \graphnode{t_2}{j_2}$ might go through intermediate nodes $\graphnode{t_3}{j_1}$, i.e., reachability is formed transitively across chains
(e.g., the path $e_6\reach e_2$ in \cref{fig:motivating-1}).
We address this challenge depending on the setting (fully dynamic or purely incremental).

On a high level, in the fully dynamic setting (\cref{subsec:ds-cssts}) in each $\dsarray{t_1}{t_2}$ we store  direct edges, i.e., 
each edge insertion is handled similarly to what was already described for $\numThreads=2$.
During queries, we perform a transitive closure operation across chains.
Taking advantage of the properties of our data structure, this transitive closure takes $O(\numThreads^3)$ time, as opposed to $O(\numEvents^2)$ time, and thus remains efficient.

In the purely incremental setting (\cref{sec:incremental_ds}), on the other hand, $\dsarray{t_1}{t_2}$ stores transitive reachability across chains, which is computed every time a new edge inserted.
Again, taking advantage of the properties of our data structure, this transitive closure takes $O(\numThreads^2)$ time, as opposed to $O(\numEvents)$ time, and is thus very efficient.
Now, queries can be performed by directly querying the respective suffix minima array, similarly to what was already described for $\numThreads=2$.

%% file: material/examples/segtree.tex
%!TEX root = ../../main.tex

\begin{figure}[h!]
	\centering
	\begin{tikzpicture}[every node/.style=draw,node distance=0.15mm, thick, level distance=0.56cm, level 1/.style={sibling distance=45mm}, level 2/.style={sibling distance=24mm}, inner sep=3, rounded corners]
		\node[](root){ \tikznodeorig{0}{3}{6} } 
		child{
			node[]{ \tikznodeorig{0}{1}{6}  } 
			child{node[]{ \tikznodeorig{0}{0}{6} } }
			child{node[]{ \tikznodeorig{1}{1}{9} } }
		}
		child{
			node[]{ \tikznodeorig{2}{3}{8}  } 
			child{node[]{ \tikznodeorig{2}{2}{8}} }
			child{node[]{ \tikznodeorig{3}{3}{10}} }
		};
	\end{tikzpicture}
	\caption{A Segment Tree representation of $A=[6, 9, 8, 10]$.}
	\label{fig:segtree}
\end{figure}

\begin{example}
Consider the Segment Tree $T$ in \cref{fig:segtree} representing the array $A=[6, 9, 8, 10]$.
We have the following.
\begin{compactitem}
	\item $\mn(A, 0) = 6$, $\mn(A, 1) = \mn(A, 2) = 8$, $\mn(A, 3) = 10$
	\item $T.\rootnode.\mn = 6$, $\node_{2, 3}.\min = 8$
	\item $\argleq(A, 7) = 0$, $\argleq(A, 9) = 2$, $\argleq(A, 11) = 3$ 
	\item $\upd(A, 3, 7)$ sets $A[3] = 7$. In $T$, this is captured by setting $\node_{2, 3}.\mn=\node_{3, 3}.\mn=7$.
\end{compactitem}
\end{example}

%% file: material/examples/subgraph.tex
%!TEX root = ../../main.tex

\input{material/examples/def}
\begin{example}
	Consider again the partial order $\partialord{\tr}$ in $\cref{fig:motivating-1}$.
	Some of the arrays $\dsarray{t_1}{t_2}$ are shown in \cref{fig:defs}.
	Observe that $\suc(e_3, 2)=e_6$ can be determined by the suffix minima query $\mn(\dsarray{1}{2}, 0)$.
	Similarly, $\pred(e_2, 1)=e_5$ can be determined by the suffix arg-minima query $\argleq(\dsarray{1}{0}, 1)$.
\end{example}

%% file: material/examples/def.tex
%!TEX root = ../../main.tex

\begin{figure}
	\centering
	\def\ystep{0.5}

	\begin{subfigure}[b]{0.10\textwidth}
		\centering
		\begin{tikzpicture}
			\setcounter{index}{0}
			\coordinate (s) at (0,0);
			\foreach \num in {$\infty$, $\infty$, $\infty$}{
				\node[minimum size=6mm, draw, rectangle] at (s) {\num};
				\node at ($(s)-(0,0.5)$) {\theindex};
				\stepcounter{index}
				\coordinate (s) at ($(s) + (0.6,0)$);
			}
		\end{tikzpicture}
		\caption{Array $\dsarray{0}{1}$.}
		\label{fig:defs-out1}
	\end{subfigure}
	~
	\begin{subfigure}[b]{0.10\textwidth}
		\centering
		\begin{tikzpicture}
			\setcounter{index}{0}
			\coordinate (s) at (0,0);
			\foreach \num in {$\infty$, $\infty$, $1$}{
				\node[minimum size=6mm, draw, rectangle] at (s) {\num};
				\node at ($(s)-(0,0.5)$) {\theindex};
				\stepcounter{index}
				\coordinate (s) at ($(s) + (0.6,0)$);
			}
		\end{tikzpicture}
		\caption{Array $\dsarray{1}{0}$.}
		\label{fig:defs-out2}
	\end{subfigure}
	~
	\begin{subfigure}[b]{0.11\textwidth}
		\centering
		\begin{tikzpicture}
			\setcounter{index}{0}
			\coordinate (s) at (0,0);
			\foreach \num in {$0$, $n$-$6$, $\infty$}{
				\node[minimum size=6mm, draw, rectangle] at (s) {\num};
				\node at ($(s)-(0,0.5)$) {\theindex};
				\stepcounter{index}
				\coordinate (s) at ($(s) + (0.65,0)$);
			}
		\end{tikzpicture}
		\caption{Array $\dsarray{1}{2}$.}
		\label{fig:defs-out3}
	\end{subfigure}
	~
	\begin{subfigure}[b]{0.10\textwidth}
		\centering
		%		\begin{tikzpicture}
			%			\setcounter{index}{0}
			%			\coordinate (s) at (0,0);
			%			\foreach \num in {$\infty$, $\infty$, $\infty$}{
				%				\node[minimum size=6mm, draw, rectangle] at (s) {\num};
				%				\node at ($(s)-(0,0.5)$) {\theindex};
				%				\stepcounter{index}
				%				\coordinate (s) at ($(s) + (0.6,0)$);
				%			}
			%		\end{tikzpicture}
		\begin{tikzpicture}
			\setcounter{index}{0}
			\coordinate (s) at (0,0);
			\foreach \num in {$\infty$}{
				\node[minimum size=6mm, draw, rectangle] at (s) {\num};
				\node at ($(s)-(0,0.5)$) {\theindex};
				\stepcounter{index}
				\coordinate (s) at ($(s) + (0.6,0)$);
			}
			\node[draw=none] at (0.6, 0) {$\ldots$};
			\setcounter{index}{7}
			\coordinate (s) at (1.2,0);
			\foreach \num in {$2$}{
				\node[minimum size=6mm, draw, rectangle] at (s) {\num};
				\node at ($(s)-(0,0.5)$) {$n$-$6$};
				\stepcounter{index}
				\coordinate (s) at ($(s) + (0.6,0)$);
			}
		\end{tikzpicture}
		\caption{Array $\dsarray{2}{1}$.}
		\label{fig:defs-out4}
	\end{subfigure}

	\caption{Representation of the partial order in \cref{fig:motivating-1}.}
	\label{fig:defs}
\end{figure}

%% file: sections/ds-sparse-dense-segment-tree.tex
\subsection{Sparse Segment Trees}\label{subsec:sparse_segment_tree}

Using $\segtree$s $T$ to solve suffix minima, each operation completes in time proportional to the height of $T$, i.e., $O(\log \numEvents)$.
In practice $\numEvents$ can be very large and thus hinder performance.
Here we develop Sparse Segment Trees ($\sparsest$s), that optimize $\segtree$s for our setting.
These are based on two main ideas, namely
(i)~\emph{minima indexing}, and
(ii)~\emph{sparse tree representation}.

Intuitively, $\segtree$s solve the more general problem of \emph{range minima queries}, where every query asks for the minima in a range $A[i: j]$, as opposed to the suffix $A[i:]$.
Minima indexing exploits the fact that our queries always concern a suffix of $A$, and allows them to often complete without traversing a full path from the root to a leaf.

Sparse tree representation is an optimization of the tree representation, which becomes more condensed the lower the density of $A$ is (i.e., the fewer non-$\infty$ entries it has), stemming from the observation that $\infty$ entries do not contribute to the queries.
In a dynamic analysis setting, $G$ normally has low cross-chain density $\CrossChainDensity$, which means that the suffix-minima arrays storing reachability across chains (as described in \cref{subsec:dynamic-suffix-minima}) are sparse.
This reduces the height of the tree, allowing both queries and updates to run faster.

\myparagraph{Minima indexing}
Each node $\node_{i, j}$ of a Sparse Segment Tree $T$ maintains a field $\pos\in [i,j]$,
that is defined recursively:~it points to the largest index of $A$ that contains the minimum element in the range $A[i: j]$,
after excluding all indexes that are stored in the $\pos$ fields of all ancestors of $\node_{i,j}$.
Formally, let $\node^1, \dots, \node^{m}$ be the ancestors of $\node_{i,j}$ in $T$.
Then
\[
\node_{i,j}.\pos= \max\left(\argmin_{\ell\in[i,j] \setminus \{\node^1.\pos,\dots, \node^{m}.\pos \} } A[\ell]\right)
\numberthis\label{eq:pos}
\]
We also set $\node_{i,j}.{\mn}=A[\node_{i,j}.\pos]$.
Observe that now $\node_{i,j}.{\mn}$ might not be the minimum value in $A[i:j]$, in contrast to the standard Segment Trees.

The intuition behind $\node_{i,j}.\pos$ is as follows.
Each query $\mn(A, i)$ starts a traversal from $T.\rootnode$ to a leaf.
If we encounter a node $\node_{i,j}$ with $\node_{i,j}.\pos\geq i$, the traversal can stop early and return that $\mn(A, i)=\node_{i,j}.\mn$.
Since we always query on a suffix of $A$, storing the \emph{largest} index pointing to a minima increases the chances of stopping early.
Moreover, we exclude the $\pos$ fields of the ancestors of $\node_{i,j}$ as these are irrelevant for $\node_{i,j}$, in the sense that this index would have already stopped the traversal in an ancestor of $\node_{i,j}$.

\input{material/examples/sst-pos}

\myparagraph{Sparse representation}
Sparse Segment Trees are tailored towards representing sparse arrays.
This allows the height of the tree to shrink, which consequently results in more efficient updates and queries.
A crucial aspect of  $\sparsest$s is that they are self-adapting and do not require any a priori information about the sparsity of the input.

Intuitively, we achieve this as follows.
Empty array indexes are never explicitly represented in the intermediate nodes of the tree.
Every intermediate node $\node$ is such that $\node.\pos$ points to a unique, non-empty entry of $A$.
The opposite is also true:~every non-empty entry of $A$ is indexed by the $\pos$ field of a node in the earliest possible level satisfying \cref{eq:pos}.

\input{material/examples/size-shrink}

One last optimization step is to flatten subtrees of small size at the leaves to array blocks.
For a node $\node_{i, j}$, if $\depth(\node_{i,j})$ is larger than a value deeming that the node represents a small sub-array of $A$, then it becomes a leaf in $T$, storing the subarray $A[i: j]$.
We call such a node $\node_{i, j}$ a \emph{block node}.
This representation is valuable when the  array contains segments which are densely populated but also far apart from each other.
%sBlock nodes were also used in other Segment Tree representations recently, such as in the $B$-ary Segment Tree~\cite{Pibiri-2021}.

\input{material/examples/block}

\input{material/algorithms/opt-ds/implementation}

\myparagraph{Operations on Sparse Segment Trees}
\cref{algo:ds-internal} implements the operations $\upd, \mn, \argleq$ on a $\sparsest$.

\SubParagraph{Function \FunctionUpdate{$\node, pos, v$}}.
This function inserts a node $\node'$ into the subtree of $\node$, with $\node'.\pos = pos$ and $\node'.\mn = v$. 
It first checks if $\node$ is initialized; if not, it creates a new node (\cref{line:ds-update-nil-check}).
If $\node$ is the root, the function checks and deletes if a node $\node''$ exists in the subtree of $\node$ where $\node''.\pos = pos$ (\cref{line:ds-update-del}).
Then, it checks if $2^{\depth(\node)} \geq \numEvents/\threshold$, where $\threshold$ is the block-size threshold. 
If so, $\node$ becomes a block node and the update is performed directly on $\node.\block$ (\cref{line:ds-update-block}).
Otherwise, it checks if  $v \leq \node.\mn$.
If so, it swaps the values of $\node.\pos$ and $\node.\mn$ with $pos$ and $v$ (\cref{line:ds-update-min-check}).
This captures that the new entry will be stored in $\node$ and the previous entry in $\node$ will be relocated to a child node.
Next, it identifies which subtree the corresponding entry belongs to, and makes a call to $\FunctionUpdateHelper$ 
(\cref{line:ds-update-left-call}, \cref{line:ds-update-right-call}).
If the subtree is not present, $\FunctionUpdateHelper$ creates a new node (\cref{line:ds-update-helper-create-node}).
Otherwise, it checks if $\pos$ is within the range of the subtree and if so it recursively calls $\upd$.
If the subtree is present but $\pos$ is not within the range,  $\FunctionCreateLowestCommonAncestor$ creates a new node whose range corresponds to the lowest common ancestor of the existing subtree and $\pos$.

\SubParagraph{Function  \FunctionMin{$\node, i$}.}
This recursive function returns $\infty$ if $\node$ is $\nil$ or $i > \node.\en$ (\cref{line:ds-min-nil-check}), which captures the end of the recursion.
Otherwise, it returns the smallest $\mn$ value in the subtree of $\node$ whose corresponding $\pos$ is greater than or equal to $i$, as follows.
First, it checks if $\node$ is a block node, in which case it returns the minimum of the block array $\node.\block$ (\cref{line:ds-min-block}).
Otherwise, if the minima indexing optimization succeeds, the traversal stops here, returning $\node.\mn$ (\cref{line:ds-min-opt}).
If both conditions fail, the function proceeds recursively on the left and right child of $\node$ %returning the minimum of the two 
(\crefrange{line:ds-min-both-call-start}{line:ds-min-end}).

\SubParagraph{Function \FunctionArgLeq{$\node, v$}.}
This function returns $-\infty$ if $\node$ is $\nil$ or $\node.\mn > v$.
Otherwise, it returns the largest $\pos$ in the subtree of $\node$ whose corresponding $\mn$ value is smaller than or equal to $v$, as follows.
It first checks if $\node$ is a block node, in which case it simply performs an array traversal on $\node.\block$ (\cref{line:ds-arqleq-block}).
Otherwise, the function uses the minima indexing information in $\node.\pos$ to check if it can return early (\cref{line:ds-argleq-optimization}).
If the check fails, it makes a recursive call on the right subtree if $\node.\rightt.\min \leq v$ (\cref{line:ds-argleq-right-call}).
If all conditions fail, it makes a recursive call on the left subtree (\cref{line:ds-argleq-left-call}).

\myparagraph{Using $\mathbftt{SSTs}$}
$\sparsest$s handle operations on the represented suffix-minima array $A$ is as follows.
Queries $\mn(A, i)$ and $\argleq(A, a)$ are handled as $\mn(T.\rootnode, i)$ and $\argleq(T.\rootnode, a)$, respectively.
On the other hand, $\upd(A, i, a)$ is handled as $\upd(T.\rootnode, i, a)$.

\myparagraph{Complexity and the height of $\mathbftt{SSTs}$}
The theoretical advantage of $\sparsest$s  over $\segtree$s is the fact that their height is bounded not only by $\log n$, but also by the density of the represented array.
This is captured in the following lemma.

\begin{restatable}{lemma}{lemsparsetreeheight}\label{lem:sparse_tree_height}
Consider an array $A$ of size $\numEvents$ with $d$ non-empty (i.e., non-$\infty$) entries, and let $T$ be its corresponding $\sparsest$.
Then the height of $T$ is bounded by $\min(\log n, d)$.
\end{restatable}

Thus, we can use \cref{lem:sparse_tree_height} to bound the time taken to perform any update and query on the represented array $A$.
In practice, however, the running time can be even smaller, as, in contrast to standard $\segtree$s, the update and query operations can stop before traversing a full path from the root to a leaf.

%% file: material/examples/sst-pos.tex
\begin{figure}[h]
	\centering
	\begin{subfigure}[t]{0.48\textwidth}
		\begin{subfigure}[b]{0.4\textwidth}
			\centering
			\scalebox{0.9}{
				\begin{tikzpicture}
					\setcounter{index}{0}
					\coordinate (s) at (0,0);
					\foreach \num in {$77$, $42$, $65$, $59$}{
						\node[minimum size=6mm, draw, rectangle] at (s) {\num};
						\node at ($(s)-(0,0.5)$) {\theindex};
						\stepcounter{index}
						\coordinate (s) at ($(s) + (0.6,0)$);
					}
					\node[draw=none] at (2.5, 0) {$\ldots$};
					\setcounter{index}{7}
					\coordinate (s) at (3.2,0);
					\foreach \num in {$100$}{
						\node[minimum size=6mm, draw, rectangle] at (s) {\num};
						\node at ($(s)-(0,0.5)$) {\theindex};
						\stepcounter{index}
						\coordinate (s) at ($(s) + (0.6,0)$);
					}
				\end{tikzpicture}
			}
			\caption{Array $A$.}
			\label{subfig:sst-pos-input}
		\end{subfigure}
		\hfill~~~~\hfill
		\begin{subfigure}[b]{0.6\textwidth}
			\centering
			\begin{tikzpicture}[every node/.style=draw,node distance=0.05mm, thick, level distance=0.7cm, level 1/.style={sibling distance=25mm}, level 2/.style={sibling distance=20mm}, inner sep=2, rounded corners]

				\node[below = 1 cm of s](root)  { \tikznodeopt{0}{7}{42}{1} }
				child{
					node[]{ \tikznodeopt{0}{3}{59}{3}  }
					child{node[]{ \tikznodeopt{0}{1}{77}{0} } }
					child{node[]{ \tikznodeopt{2}{3}{65}{2} } }
				}
				child{
					node[draw=none]{\ldots}
				};
			\end{tikzpicture}
			\caption{$\segtree$ representation of $\dsarray{t_1}{t_2}$.}
			\label{subfig:sst-pos-ds}
		\end{subfigure}
	\end{subfigure}
	\caption{$\segtree$ representation with minima indexing.
 The entries $A[4\colon6]$ are $> 59$. The notation $\node_{i, j}\colon (x, y)$ denotes $\node_{i, j}.\mn = x$, $\node_{i, j}.\pos = y$.
	}
	\label{fig:sst-pos}
\end{figure}

\begin{example}
	Consider the array $A$ in \cref{subfig:sst-pos-input}, and part of the corresponding $\segtree$ in \cref{subfig:sst-pos-ds}.
	We have $\node_{0, 7}.\pos=1=\arg\min_{i}A[i]$.
	A query $\mn(A, 0)$ is now directly answered at the root, as $\node_{0, 7}.\pos \in [0:7]$.
	On the other hand, the query $\mn(A, 2)$ is not answered at the root as $\node_{0, 7}.\pos \not \in [2:7]$.
	The procedure then traverses its child $\node_{0, 3}$.
	We have $\node_{0, 3}.\pos = 3$, which indexes the smallest element in $A[0 : 3]$ after excluding $\node_{0, 7}.\pos´$.
	The traversal stops here, as $\node_{0, 3}.\pos \in [2:7]$.
\end{example}

%% file: material/examples/size-shrink.tex
%!TEX root = ../../main.tex

\begin{figure}[t]
	\centering
	\begin{subfigure}[t]{0.48\textwidth}        
		\begin{subfigure}[b]{0.4\textwidth}        
			\centering
			\scalebox{0.9}{
				\begin{tikzpicture}
					\setcounter{index}{0}
					\coordinate (s) at (0,0);
					\foreach \num in {$\infty$, $\infty$, $65$, $\infty$}{
						\node[minimum size=6mm, draw, rectangle] at (s) {\num};
						\node at ($(s)-(0,0.5)$) {\theindex};
						\stepcounter{index}
						\coordinate (s) at ($(s) + (0.6,0)$);
					}
					\node[draw=none] at (2.5, 0) {$\ldots$};
					\setcounter{index}{7}
					\coordinate (s) at (3.2,0);
					\foreach \num in {$\infty$}{
						\node[minimum size=6mm, draw, rectangle] at (s) {\num};
						\node at ($(s)-(0,0.5)$) {\theindex};
						\stepcounter{index}
						\coordinate (s) at ($(s) + (0.6,0)$);
					}
				\end{tikzpicture}
			}
			\caption{Input Array $A_1$.}
			\label{fig:ex4-input-1}
		\end{subfigure}
		\begin{subfigure}[b]{0.6\textwidth}
			\centering    
			\begin{tikzpicture}[every node/.style=draw,node distance=5mm, thick, level 1/.style={sibling distance=75mm}, level 2/.style={sibling distance=55mm}, inner sep=2, rounded corners]
				\node[](root){ \tikznodeopt{0}{7}{65}{2} };
				%child[missing]{}
				%child{
					%    node[]{ \tikznode{1}{1}{3}{4}{60}{3}  }
					%child{node[]{ \tikznode{0}{2}{3}{3}{2}{3} } }
					%child{node[]{ \tikznode{0}{3}{4}{4}{1}{4} } }
					%    };
			\end{tikzpicture}
			\addtocounter{subfigure}{3}\caption{Sparse $\segtree$ representation of $A_1$.}
			\label{fig:ex4-ds-1}
		\end{subfigure}
	\end{subfigure}
	\\
	\begin{subfigure}[t]{0.48\textwidth}
		\begin{subfigure}[t]{0.4\textwidth}        
			\centering
			\scalebox{0.9}{
				\begin{tikzpicture}
					\setcounter{index}{0}
					\coordinate (s) at (0,0);
					\foreach \num in {$\infty$, $\infty$, $65$, $42$}{
						\node[minimum size=6mm, draw, rectangle] at (s) {\num};
						\node at ($(s)-(0,0.5)$) {\theindex};
						\stepcounter{index}
						\coordinate (s) at ($(s) + (0.6,0)$);
					}
					\node[draw=none] at (2.5, 0) {$\ldots$};
					\setcounter{index}{7}
					\coordinate (s) at (3.2,0);
					\foreach \num in {$\infty$}{
						\node[minimum size=6mm, draw, rectangle] at (s) {\num};
						\node at ($(s)-(0,0.5)$) {\theindex};
						\stepcounter{index}
						\coordinate (s) at ($(s) + (0.6,0)$);
					}
				\end{tikzpicture}
			}
			\addtocounter{subfigure}{-4}\caption{Input Array $A_2$.}
			\label{fig:ex4-input-2}
		\end{subfigure}
		\hfill~~\hfill
		\begin{subfigure}[t]{0.6\textwidth} 
			\centering      
			\begin{tikzpicture}[every node/.style=draw,node distance=0.2mm, thick, level distance=0.7cm, level 1/.style={sibling distance=30mm}, level 2/.style={sibling distance=40mm}, inner sep=2, rounded corners]
				\node[below = 1 cm of s](root){ \tikznodeopt{0}{7}{42}{3} } 
				child{
					node[]{ \tikznodeopt{2}{2}{65}{2}  }
				}
				child[missing]{};
			\end{tikzpicture}
			\addtocounter{subfigure}{3}\caption{Sparse $\segtree$ representation of $A_2$.}
			\label{fig:ex4-ds-2}
		\end{subfigure}
	\end{subfigure}		
	\\
	\begin{subfigure}[t]{0.48\textwidth}
		\begin{subfigure}[t]{0.4\textwidth}        
			\centering
			\scalebox{0.9}{
				\begin{tikzpicture}
					\setcounter{index}{0}
					\coordinate (s) at (0,0);
					\foreach \num in {$59$, $\infty$, $65$, $42$}{
						\node[minimum size=6mm, draw, rectangle] at (s) {\num};
						\node at ($(s)-(0,0.5)$) {\theindex};
						\stepcounter{index}
						\coordinate (s) at ($(s) + (0.6,0)$);
					}
					\node[draw=none] at (2.5, 0) {$\ldots$};
					\setcounter{index}{7}
					\coordinate (s) at (3.2,0);
					\foreach \num in {$\infty$}{
						\node[minimum size=6mm, draw, rectangle] at (s) {\num};
						\node at ($(s)-(0,0.5)$) {\theindex};
						\stepcounter{index}
						\coordinate (s) at ($(s) + (0.6,0)$);
					}
				\end{tikzpicture}
			}
			\addtocounter{subfigure}{-4}\caption{Input Array $A_3$.}
			\label{fig:ex4-input-3}
		\end{subfigure}
		\hfill~~\hfill
		\begin{subfigure}[t]{0.6\textwidth}
			\centering    
			\begin{tikzpicture}[every node/.style=draw,node distance=0.05mm, thick, level distance=0.7cm, level 1/.style={sibling distance=30mm}, level 2/.style={sibling distance=26mm}, inner sep=2, rounded corners]
				
				\node[below = 1 cm of s](root)  { \tikznodeopt{0}{7}{42}{3} } 
				child{
					node[]{ \tikznodeopt{0}{3}{59}{0}  }
					child{edge from parent[draw=none]}
					child{node[]{ \tikznodeopt{2}{2}{65}{2} } }
				}
				child[missing]{};
			\end{tikzpicture}
			\addtocounter{subfigure}{3}\caption{Sparse $\segtree$ representation of $A_3$.}
			\label{fig:ex4-ds-3}
		\end{subfigure}
	\end{subfigure}
	\\
	\begin{subfigure}[t]{0.48\textwidth}
		\begin{subfigure}[t]{0.4\textwidth}        
			\centering
			\scalebox{0.9}{
				\begin{tikzpicture}
					\setcounter{index}{0}
					\coordinate (s) at (0,0);
					\foreach \num in {$59$, $\infty$, $65$, $42$}{
						\node[minimum size=6mm, draw, rectangle] at (s) {\num};
						\node at ($(s)-(0,0.5)$) {\theindex};
						\stepcounter{index}
						\coordinate (s) at ($(s) + (0.6,0)$);
					}
					\node[draw=none] at (2.5, 0) {$\ldots$};
					\setcounter{index}{7}
					\coordinate (s) at (3.2,0);
					\foreach \num in {$13$}{
						\node[minimum size=6mm, draw, rectangle] at (s) {\num};
						\node at ($(s)-(0,0.5)$) {\theindex};
						\stepcounter{index}
						\coordinate (s) at ($(s) + (0.6,0)$);
					}
				\end{tikzpicture}
			}
			\addtocounter{subfigure}{-4}\caption{Input Array $A_4$.}
			\label{fig:ex4-input-4}
		\end{subfigure}
		\hfill~~\hfill
		\begin{subfigure}[t]{0.6\textwidth}
			\centering
			\begin{tikzpicture}[every node/.style=draw,node distance=0.2mm, thick, level distance=0.7cm, level 1/.style={sibling distance=25mm}, level 2/.style={sibling distance=20mm}, inner sep=2, rounded corners]
				
				\node[below = 1 cm of s](root)  { \tikznodeopt{0}{7}{13}{7} } 
				child{
					node[]{ \tikznodeopt{0}{3}{42}{3}  }
					child{node[]{ \tikznodeopt{0}{0}{59}{0} } }
					child{node[]{ \tikznodeopt{2}{2}{65}{2} } }
				}
				child[missing]{};
			\end{tikzpicture}
			\addtocounter{subfigure}{3}\caption{Sparse $\segtree$ representation of $A_4$.}
			\label{fig:ex4-ds-4}
		\end{subfigure}
	\end{subfigure}
	\caption{Sparse $\segtree$ representation. The notation $\node_{i, j}\colon (x, y)$ denotes $\node_{i, j}.\mn = x$, $\node_{i, j}.\pos = y$.
	}
	\label{fig:ex4}
\end{figure}

\begin{example}
	Consider an array $A$ of length $8$, initially empty.
	\crefrange{fig:ex4-input-1}{fig:ex4-input-4} sequentially update $A$ with new entries, and \crefrange{fig:ex4-ds-1}{fig:ex4-ds-4} show the evolution of the corresponding $\sparsest$ $T$.
	In \cref{fig:ex4-ds-1}, the unique node of $T$ represents that $A$ has only one non-empty entry, at index $2$ with value $65$.
	In contrast, a standard (non-sparse) Segment Tree would contain $15$ nodes and have height $3$.
	\cref{fig:ex4-ds-2} shows the update of $T$ after setting $A[3]=42$.
	Since $42$ is now the smallest value of $A$, it is stored in the root,
  while a new node $\node_{2, 2}$ becomes the left child of the root, with its $\pos$ and $\mn$ fields set according to \cref{eq:pos}.
	Again, in the standard $\segtree$ representation, $\node_{2,2}$ would have nodes $\node_{0, 3}$ and $\node_{2, 3}$ as ancestors, but these nodes are never created in our $\sparsest$.
	\cref{fig:ex4-ds-3}, shows another update of $T$ after setting $A[0]=59$.
	As $59\leq \node_{0,7}.\mn$, the root remains intact. 
	The index $\,0$ belongs to the left subtree of the root and $59<\node_{2,2}.\mn$.
	This causes the new entry to be inserted to a fresh node $\node_{0, 3}$ and assigned as the left child of the root, taking the existing node $\node_{2,2}$ as its right child.
	\cref{fig:ex4-ds-4} shows the final update of $T$ after setting $A[7]=13$.
	Since $13$ is now the smallest value of $A$, it is stored in the root,
	while the previous value stored in the root is recursively pushed downwards.
\end{example}

%% file: material/examples/block.tex
%!TEX root = ../../main.tex

\begin{figure}[h!]
	\centering
	\begin{subfigure}[t]{0.40\textwidth}        
		\begin{tikzpicture}[every node/.style=draw,node distance=0.2mm, thick, level distance=0.7cm, level 1/.style={sibling distance=50mm}, level 2/.style={sibling distance=26mm}, inner sep=2, rounded corners]
			\node[below = 1 cm of s](root)  { \tikznodeopt{0}{64}{10}{33} } 
			child{
				node[]{ \tikznodeopt{1}{1}{50}{1}  }
			}
			child{
				node[draw=none]{\ldots}
				child{
					node[]{\tikznodeopt{32}{39}{10}{33} } 
					child{
						}
					}
				child{edge from parent[draw=none]}
			};
		\end{tikzpicture}
	\end{subfigure}
	\\\vspace{-0.05cm}
	\begin{subfigure}[t]{0.15\textwidth} 
		\scalebox{0.85}{
			\begin{tikzpicture}
				\node[draw=none] at (-0.7, 0) {$A:$};
				\setcounter{index}{0}
				\coordinate (s) at (0,0);
				\foreach \num in {$\infty$, $50$, $\infty$, $\infty$}{
					\node[minimum size=6mm, draw, rectangle] at (s) {\num};
					\node at ($(s)-(0,0.5)$) {\theindex};
					\stepcounter{index}
					\coordinate (s) at ($(s) + (0.6,0)$);
				}
				\node[draw=none] at (2.5, 0) {$\ldots$};
				\setcounter{index}{32}
				\coordinate (s) at (3.2,0);
				\foreach \num in {$11$, $10$, $15$, $\infty$, $13$, $22$, $24$, $29$}{
					\node[minimum size=6mm, draw, rectangle] at (s) {\num};
					\node at ($(s)-(0,0.5)$) {\theindex};
					\stepcounter{index}
					\coordinate (s) at ($(s) + (0.6,0)$);
				}
				\node[draw=none] at (8.1, 0) {$\ldots$};
			\end{tikzpicture}
		}
	\end{subfigure}
	\hfill~\hfill
	\caption{Flatting small subtrees to block nodes.}
	\label{fig:ex-block}
\end{figure}
\begin{example}
Consider the array $A$ and its corresponding $\sparsest$ in \cref{fig:ex-block}. 
$A$ contains only a single entry in the range $0,\ldots,31$.
This entry is sparsely represented in the left subtree of the root.
On the other hand, $A$ is dense in the range $32,\ldots,39$, which would require a full subtree.
Instead, when the range of the represented subarray is smaller than a threshold $\threshold$,
the corresponding subtree is flattened to an array block.
\end{example}

%% file: material/algorithms/opt-ds/implementation.tex
%!TEX root = ../../../main.tex
%\setlength{\textfloatsep}{12pt}

\begin{algorithm*}[ht!]
	\small
	\DontPrintSemicolon
	\SetInd{0.28em}{0.35em}
	\BlankLine
	\vspace*{-\multicolsep}
	%\TODO{Show how the transition to blocks is made}
	\begin{multicols}{2}
		
		%		\MyFunction{\FunctionMin{$\node, \start, \en$}}{
			%			\label{line:ds-min-start}
			%			\lIf{$\node = \nil$}{\Return $\infty$ \label{line:ds-min-nil-check}}
			%			\lIf {$\node.\pos \in [\start, \en]$} {
				%				\Return $\node.\mn$; \label{line:ds-min-opt}
				%			} \uElse {
				%				$\midd = \node.\start + (\node.\en - \node.\start) / 2$\\
				%				\lIf {$\en \leq \midd$}{\Return $\node.\leftt, \start, \en)$ \label{line:ds-min-left-call}}
				%				\lElseIf{$\start > \midd$} {
					%					\Return $\mn(\node.\rightt, \start, \en)$ \label{line:ds-min-right-call}
					%				} 
				%				
				%				\uElse{
					%					%Implicity assume that if nd.left = $\infty$ then min returns $\infty$
					%					\label{line:ds-min-both-call-start}
					%					$l \xleftarrow[]{} \mn(\node.\leftt, \start, \midd)$\\
					%					$r \xleftarrow[]{} \mn(\node.\rightt, \midd+1, \en)$\\
					%					\lIfElse{$l < r$}{\Return $l$}{\Return $r$} \label{line:ds-min-end}
					%				}
				%			}
			%		}

		\MyFunction{\FunctionUpdate{$\node, pos, v$}} { 
			\label{line:ds-update-start}
			
			\lIf {$\node = \nil$} {
				$\node \gets \FunctionCreateNode(pos, v)$
				\label{line:ds-update-nil-check}
			} 
			\lElseIf{$\node = T.\rootnode$} {
				$\node \gets \delnode(\node, pos)$\label{line:ds-update-del}
				%$\upd(\node, pos, v)$\label{line:ds-update-del-update}
			}			
			\lElseIf {$2^{\depth(\node)} \geq \numEvents/\threshold$} {
				$\node.\block[pos] = v$; \label{line:ds-update-block}
			} 
			
%			\lElseIf {$\node.\isleaf$} {
%				$\node.\mn \gets v$
%				\label{line:ds-update-leaf}
%				%Implicitly assume that v is at least as large as \node.\mn.
%			} 
%			\uElseIf{$\node.\pos = \pos$} {
%				$\node \gets \delnode(\node, pos)$\label{line:ds-update-del}\\
%				$\upd(\node, pos, v)$\label{line:ds-update-del-update}
%			}
			\uElse {
				\uIf {$v < \node.\mn \lor (v = \node.\mn \land pos > \node.\pos)$} {
					$\node.\mn, \node.\pos, v, pos \gets v, pos, \node.\mn, \node.\pos$
					\label{line:ds-update-min-check}
				}
				Let~$mid \gets \node.\start + \lfloor (\node.\en - \node.\start) / 2 \rfloor$\\
				\uIf {$pos \leq mid$} { \label{line:ds-update-pos-check-1}
					%$\node.\leftt \gets \hspace{1cm} \FunctionUpdateHelper(\node.\leftt, \node.\start, mid, pos, v)$
					%			\uIf {$\node.\leftt \neq \nil$} {
						%				$\upd(\node.\leftt, pos, v)$
						%				\label{line:ds-update-left-1}
						%			} 
					%			\lElse {
						%				$\node.\leftt \gets \FunctionCreateNode(\node.start, mid, pos, v)$
						%				\label{line:ds-update-left-2}
						%			}
					
					$\node.\leftt \gets \FunctionUpdateHelper(\node.\leftt, pos, v)$\label{line:ds-update-left-call}
					
				} \uElse{ \label{line:ds-update-pos-check-2}
					%			\uIf {$\node.\rightt \neq \nil$} {
						%				$\upd(\node.\rightt, pos, v)$
						%				\label{line:ds-update-right-1}
						%			} 
					%			\lElse {
						%				$\node.\rightt \gets \FunctionCreateNode(mid+1, \node.\en, pos, v)$
						%				\label{line:ds-update-right-2}
						%			}
					$\node.\rightt \gets \FunctionUpdateHelper(\node.\rightt, pos, v)$
					%$\node.\rightt \gets \FunctionUpdateHelper(\node.\rightt, mid+1, \node.\en, pos, v)$
					\label{line:ds-update-right-call}
				}
			}\label{line:ds-update-end}
		}
		%		\uIf {$pos \leq mid$} {
			%			\lIf {$\node.\leftt = \nil$} {
				%				$\node.\leftt \gets createNode(\node.\start, mid, pos, v)$
				%			} \uElse {
				%				\lIf {$pos \in [\node.\leftt.\start, \node.\leftt.\en]$} {
					%					$\upd(\node.\leftt, pos, v)$
					%				}
				%				\lElse {
					%					$\node.\leftt \gets createNodes(\node.\leftt, \node.\start, mid, pos, v)$}
				%			}
			%		} 
		%		\uElse {
			%			\lIf {$\node.\rightt = \nil$} {
				%				$\node.\rightt \gets createNode(mid+1, \node.\en, pos, v)$
				%			} \uElse {
				%				\lIf {$pos \in [\node.\rightt.\start, \node.\rightt.\en]$}{
					%					$\upd(\node.\rightt, pos, v)$
					%				}
				%				\lElse {
					%					$\node.\rightt \gets createNodes(\node.\rightt, mid+1, \node.\en, pos, v)$}
				%			}
			%		}
		%	\BlankLine
		%	\BlankLine
		\MyFunction{\FunctionUpdateHelper{$\node, pos, v$}}{ 
			\lIf {$\node = \nil$} {
				$\node \gets \FunctionCreateNode(pos, v)$
				\label{line:ds-update-helper-create-node}
			} 
			\lElseIf {$pos \in [\node.\start, \node.\en]$}{
				$\upd(\node, pos, v)$
			}
			\lElse {
				$\node \gets \FunctionCreateLowestCommonAncestor(\node, pos, v)$}
			\Return{$\node$}
		}

		\BlankLine
		\BlankLine
		\MyFunction{\FunctionMin{$\node, i$}}{
			\label{line:ds-min-start}
			\lIf{$\node = \nil \lor i > \node.\en$}{\Return $\infty$ \label{line:ds-min-nil-check}}
			\lIf {$\node.\block \neq \nil$} {
				\Return $\min(\node.\block, i)$; \label{line:ds-min-block}
			} 
			
			\lIf {$\node.\pos \geq i$} {
				\Return $\node.\mn$; \label{line:ds-min-opt}
			} 
			
			\uElse {
				%$\midd = \node.\start + (\node.\en - \node.\start) / 2$\\
				%\lIf {$\en \leq \midd$}{\Return $\mn(\node.\leftt, \start, \en)$ \label{line:ds-min-left-call}}
				%\lElseIf{$\start > \midd$} {
					%	\Return $\mn(\node.\rightt, \start, \en)$ \label{line:ds-min-right-call}
					%} 
				
				%\uElse{
					%Implicity assume that if nd.left = $\infty$ then min returns $\infty$
					\label{line:ds-min-both-call-start}
					$l \xleftarrow[]{} \min(\node.\leftt, i)$\\
					$r \xleftarrow[]{} \min(\node.\rightt, i)$\\
					\lIfElse{$l < r$}{\Return $l$}{\Return $r$} \label{line:ds-min-end}
					%}
			}
		}
		
		\BlankLine
		\BlankLine
		\MyFunction{\FunctionArgLeq{$\node, v$}}{ \label{line:ds-argleq-start}
			\lIf{$\node = \nil \lor \node.\mn > v$} {
				\Return $-\infty$ \label{line:ds-argleq-check-undefined} %$\mathtt{undefined}$ 
			}
			\lIf {$\node.\block \neq \nil$} {
				\Return $\argleq(\node.\block, v)$; \label{line:ds-arqleq-block}
			} 
			%Let $\node_{i, j}, \node_{i', j'}\gets \node.\leftt, \node.\rightt$ \\
			
			%\lIf{$\node.\mn = v \lor \node.\pos \geq \node.\rightt.\en \lor \node.\isleaf$} {
				\lIf{$\node.\pos \geq \node.\leftt.\en \land \node.\pos \geq \node.\rightt.\en$} {
					\Return $\node.\pos$ \label{line:ds-argleq-optimization}
				}
				
				\lElseIf{$\node.\rightt.\mn \leq v$} {
					\Return $max(\node.\pos, \argleq(\node.\rightt, v))$\hspace{-0.1cm} \label{line:ds-argleq-right-call}
				} 
%\lElseIf{$\node.\pos \geq \node.\leftt.\en$} {
%					\Return $\node.\pos$ \label{line:ds-argleq-optimization-2} % This corresponds to the case where we don't have a right child but the pos of the node belongs to the range in the right child of the node 
%				}  
				\lElse {\Return $max(\node.\pos, \argleq(\node.\leftt, v))$}
				\label{line:ds-argleq-end} \label{line:ds-argleq-left-call}
				
			}

		\end{multicols}
		\BlankLine
		
		\caption{
			The Sparse Segment Tree.
		}
		\label{algo:ds-internal}
		\vspace*{-\multicolsep}
	\end{algorithm*}
	\normalsize

%% file: sections/ds-cssts.tex
\subsection{Collective Sparse Segment Trees}\label{subsec:ds-cssts}

We are now ready to describe $\ds$, leading to \cref{thm:ds_dynamic}.

\input{material/algorithms/opt-ds/interface-full}

\myparagraph{$\mathbftt{CSSTs}$}
For every two distinct chains $t_1, t_2\in [\numThreads]$, $\ds$ maintain a suffix-minima array $\dsarray{t_1}{t_2}$ storing direct edges from $t_1$ to $t_2$, together with a min-heap $\edgeheap_{t_1}^{t_2}[j_1]$, for each $j_1\in [\numEvents]$, as described in \cref{subsec:dynamic-suffix-minima}.
Suffix-minima queries on $\dsarray{t_1}{t_2}$ are handled using a $\sparsest$, as described in \cref{subsec:sparse_segment_tree}.
The pseudocode is shown in \cref{algo:ds-full-interface}.
The functions $\FunctionInsertEdge$ and $\FunctionDeleteEdge$ follow directly the description in \cref{subsec:dynamic-suffix-minima}, i.e., they manipulate the arrays $\dsarray{t_1}{t_2}$ and min-heaps $\edgeheap_{t_1}^{t_2}[j_1]$.

The function  $\FunctionQuery$ is implemented via a $\suc$ query, which is more involved.
As we have $\numThreads>2$ chains, $\suc$ runs a forward traversal from $\graphnode{t_1}{j_1}$ to discover transitive reachability across chains.
This is done by repeated $\mn$ queries in the suffix minima arrays $\dsarray{t'_2}{t'_1}$ (\cref{line:ds-full-v}), which find the earliest node of $t'_1$ that has an edge from the currently earliest node of $t'_2$ found to be reachable from $\graphnode{t_1}{j_1}$.
Crucially, this computation completes in $O(\numThreads^3)$ steps, as opposed to $O(\numEvents^2)$ steps normally required for a graph traversal.

\input{material/examples/ds-full}

\myparagraph{Analysis}
We now state the properties of fully dynamic $\ds$.
First, since the arrays $\dsarray{t_1}{t_2}$ only store direct edges, the following sparsity property is straightforward.

\begin{restatable}[Sparsity]{lemma}{lemdssparsity}\label{lem:ds-sparsity}
Let $d$ be the cross-chain density of $G$.
Then the density of each array $\dsarray{t_1}{t_2}$ is bounded by $d$.
\end{restatable}

The use of min-heaps guarantees the following invariant.

\begin{restatable}[Invariant]{lemma}{lemdsinvariant}\label{lem:ds-invariant}
For every distinct $t_1, t_2\in [\numThreads]$,
we have $\dsarray{t_1}{t_2}[j_1] = \min(\{j_2\in [\numEvents].~\graphnode{t_1}{j_1}\to \graphnode{t_2}{j_2}\})$.
\end{restatable}

A \emph{crossing path} is a sequence of nodes $\pi\colon \graphnode{t_0}{j_0},\dots, \graphnode{t_{m-1}}{j_{m-1}}$ such that for each $\ell\in[m-2]$ we have $t_{\ell}\neq t_{\ell+1}$ and there are $i_1,i_2\in [\numEvents]$ such that
(i)~$i_1\geq j_{\ell}$,
(ii)~$i_2\leq j_{\ell+1}$, and
(iii)~$\graphnode{t_{\ell}}{i_1}\to \graphnode{t_{\ell+1}}{i_2}$.
Clearly, for any two nodes $\graphnode{t_1}{j_1}$ and $\graphnode{t_2}{j_2}$, if $\graphnode{t_1}{j_1}\reach \graphnode{t_2}{j_2}$, then there exists a crossing path between them of length at most $\numThreads$.
The following lemma implies the correctness of queries.

\begin{restatable}{lemma}{lemdscomplexitysuc}\label{lem:ds-complexity-suc}
For every $i\in [\numThreads]$, after the $i$-th iteration of the do-while loop in \cref{line:algo_ds_do},
$\closure[t'_1]$ points to the earliest node in chain $t'_1$ that is reachable from $\graphnode{t_1}{j_1}$ via a crossing path of length $\leq i+1$.
%For every $i\in[\numThreads]$, the $i$-th iteration 
%there are $i$ threads $t$ for which $\closure_{t_1}^{t}[j_1]=\min(\{j\in [\numEvents].~\graphnode{t_1}{j_1}\reach \graphnode{t}{j}\})$.
\end{restatable}

Regarding the time complexity, due to \cref{lem:ds-sparsity} and \cref{lem:sparse_tree_height}, each operation in a suffix minima array $\dsarray{t_1}{t_2}$ takes $O(\min(\log \numEvents, \CrossChainDensity))$ time.
Since the maximum out-degree is $\MaxDegree$, each operation on a min heap $\edgeheap_{t_1}^{t_2}[j_1]$ runs in $O(\log \MaxDegree)$ time.
Thus, updates take  $O(\max(\log \MaxDegree,\min(\log \numEvents,\CrossChainDensity)))$ time, while, due to \cref{lem:ds-complexity-suc},
queries take $O(\numThreads^3 \min(\log \numEvents, \CrossChainDensity))$ time, arriving at \cref{thm:ds_dynamic}.

\myparagraph{Space usage}
We now discuss the space usage of $\ds$.
Consider a chain DAG $G$ of $\numEvents$ nodes, $\numThreads$ chains, and cross-chain density $\CrossChainDensity$.
In this fully dynamic setting, $\ds$ store all edges in the min heaps $\edgeheap_{t_1}^{t_2}[j_1]$, as edge deletions requires to remember all edges added so far.
The suffix minima arrays $\dsarray{t_1}{t_2}$ store in total $O(\CrossChainDensity \numThreads)$ entries, as 
(i)~each chain has at most $\CrossChainDensity$ entries that have at least one successor to another chain, and
(ii)~there are $<\numThreads$ other chains.
Naturally, $\CrossChainDensity\leq \numEvents$, thus in the worst case the above expression becomes $O(\numEvents \numThreads)$.
This is the same space complexity as Vector Clocks.
However, Vector Clocks do not exploit the sparsity of $G$.
As we will see in \cref{sec:experiments}, in practice typically $\CrossChainDensity \Lt \numEvents$, which makes $\ds$ more space-efficient than Vector Clocks.

%% file: material/algorithms/opt-ds/interface-full.tex
%!TEX root = ../../../main.tex
%\setlength{\textfloatsep}{10pt}
\begin{algorithm}[ht!]
	\small
	\DontPrintSemicolon
	\SetInd{0.3em}{0.3em}
		
	\BlankLine
	\MyFunction{\FunctionSucc{$\graphnode{t_1}{j_1}, t_2$}}{\label{line:ds-full-succ-1}
		\ForEach{$t_{1}' \in [k] \setminus \set{t_1}$} {\label{line:ds-full_foreach}
			$\closure[t_1'] \gets \min(\dsarray{t_1}{t_{1}'}, j_1)$ %$\suc(\graphnode{i_1}{j_1}, i_{1}')$
		}

		\SetKwRepeat{Do}{do}{while}
		\Do{$\changed$} {\label{line:algo_ds_do}
			$\changed \xleftarrow[]{} \mathbf{false}$\\
			\ForEach{$t_{1}' \in [k] \setminus \set{t_1}$}{
				\ForEach{$t_{2}' \in [k] \setminus \set{t_1}$}{
					\uIf{$t_{1}' \neq t_{2}'$}{
						Let $v \gets \min(\dsarray{t_{2}'}{t_{1}'}, \closure[t_{2}'])$\label{line:ds-full-v}\\
						%\suc(\graphnode{i_{2}'}{\closure_{ i_1}^{i_{2}'}[j_1]}, i_{1}')$\\
						\uIf{$v < \closure[t_{1}']$} {
							$\closure[t_{1}'] \gets v$\label{line:ds-full-update_closure}\\
							$\changed \xleftarrow[]{} \mathbf{true}$
						}
					}
				}
			}
			
		}
		
		\Return $\closure[t_2]$
%		\lIfElse{$\closure_{i_1}^{i_2}[j_1] \leq j_2$}{
%			\Return True
%		} {
%			\Return False
%		}
	}

	\BlankLine
	\BlankLine
	\MyFunction{\FunctionPred{$\graphnode{t_1}{ j_1}, t_2$}}{\label{line:ds-full-pred-1}
		\tcp{Similar to successor}
		%\Return $\argleq(\dsarray{t_2}{t_1}, j_1)$
	}
	
	\BlankLine
	\BlankLine
	\MyFunction{\FunctionQuery{$\graphnode{t_1}{ j_1}, \graphnode{t_2}{ j_2}$}}{\label{line:ds-full-query-1}
		\lIfElse{$\suc(\graphnode{ t_1}{ j_1}, t_2) \leq j_2$}{\Return True}{\Return False}
	}
	
	\BlankLine
	\BlankLine
	\MyFunction{\FunctionInsertEdge{$\graphnode{t_1}{j_1}, \graphnode{t_2}{j_2}$}}{
		\lIf{$j_2 < \edgeheap_{t_1}^{t_2}[j_1].\mn$}{
			$\upd(\dsarray{t_{1}}{t_{2}}, j_{1}, j_{2})$\label{line:ds-full-interface-ins-1}
		}
		$\edgeheap_{t_1}^{t_2}[j_1].\insertitem{(j_2)}$
	}
	
	\BlankLine
	\BlankLine
	\MyFunction{\FunctionDeleteEdge{$\graphnode{t_1}{j_1}, \graphnode{t_2}{j_2}$}} {
		
		\uIf{$j_2 = \edgeheap_{t_1}^{t_2}[j_1].\mn$} {
			\label{line:ds-full-interface-1}
			$\edgeheap_{t_1}^{t_2}[j_1].\del{(j_2)}$\label{line:ds-full-interface-2}\\
			$\upd(\dsarray{t_{1}}{t_{2}}, j_{1}, \edgeheap_{t_1}^{t_2}[j_1].\mn)$\label{line:ds-full-interface-3}
		} \lElse{
			$\edgeheap_{t_1}^{t_2}[j_1].\del{(j_2)}$\label{line:ds-full-interface-4}
		}
	}

	\caption{
		Fully dynamic $\ds$.
	}
	\label{algo:ds-full-interface}
\end{algorithm}
\normalsize

%% file: material/examples/ds-full.tex
%!TEX root = ../../main.tex

\begin{figure}%[h!]
	\centering
	\def\ystep{0.8}
	\begin{subfigure}[t]{0.45\textwidth}
		\centering
		\scalebox{0.95}{
			\definecolor{offwhite}{HTML}{000000}
			\tikzset{
				> = stealth,
				every path/.append style = {
					arrows = ->,
				},
			}
			\tikz{
				\node (a) at (0,-0.3) {Thread $0$};
				\node (a0) at (0,-1*\ystep) {$\graphnode{0}{0}$};
				\node (a1) at (0,-2*\ystep) {$\graphnode{0}{1}$};
				\node (a2) at (0,-3*\ystep) {$\graphnode{0}{2}$};
				\path (a0) edge (a1);
				\path (a1) edge (a2);
				
				\node (b) at (2,-0.3) {Thread $1$};
				\node (b0) at (2,-1*\ystep) {$\graphnode{1}{0}$};
				\node (b1) at (2,-2*\ystep) {$\graphnode{1}{1}$};
				\node (b2) at (2,-3*\ystep) {$\graphnode{1}{2}$};
				\path (b0) edge (b1);
				\path (b1) edge (b2);
				
				\node (c) at (4,-0.3) {Thread $2$};
				\node (c0) at (4,-1*\ystep) {$\graphnode{2}{0}$};
				\node (c1) at (4,-2*\ystep) {$\graphnode{2}{1}$};
				\node (c2) at (4,-3*\ystep) {$\graphnode{2}{2}$};
				\path (c0) edge (c1);
				\path (c1) edge (c2);
				
				\node (d) at (6,-0.3) {Thread $3$};
				\node (d0) at (6,-1*\ystep) {$\graphnode{3}{0}$};
				\node (d1) at (6,-2*\ystep) {$\graphnode{3}{1}$};
				\node (d2) at (6,-3*\ystep) {$\graphnode{3}{2}$};
				\path (d0) edge (d1);
				\path (d1) edge (d2);
				
				%\path (a2) edge (d2);
				%\path (b1) edge (c1);
				%\path (c1) edge (d1);
				\path (c2) edge (d1);

				\draw[mosolid] (a1) to node[pos=0.4, above]{$\color{bred}{\circled{1}}$} (b0);
				\draw[mosolid, bend right=15] (a2) to node[pos=0.5, above]{$\color{bred}{\circled{2}}$} (d2);
								
				\draw[mosolid] (b1) to node[pos=0.4, above]{$\color{bred}{\circled{3}}$} (c1);
				
				\draw[mosolid] (c2) to node[pos=0.4, above]{$\color{bred}{\circled{4}}$} (d1);
				%\draw[mosolid] (b1) to (c0);
				%\draw[po, bend right=15] (a2) to (d2);
				%\node[hidden] (u) at (1.5,2) {$U$};
			}
		} 
		%\caption{A chain DAG $G$.}
		\label{fig:ds-full-dag}
	\end{subfigure}
	\\
%	\begin{subfigure}[t]{0.14\textwidth}        
%		\centering
%		\begin{tikzpicture}
%			\setcounter{index}{0}
%			\coordinate (s) at (0,0);
%			\foreach \num in {$\infty$, $0$, $\infty$}{
%				\node[minimum size=6mm, draw, rectangle] at (s) {\num};
%				\node at ($(s)-(0,0.5)$) {\theindex};
%				\stepcounter{index}
%				\coordinate (s) at ($(s) + (0.6,0)$);
%			}
%		\end{tikzpicture}
%		%\caption{Leaf nodes of $\segtree_{1}^{2}$.}
%		\caption{Array $\dsarray{0}{1}$.}
%		\label{fig:interface-out1}
%	\end{subfigure}
%	~~
%	\begin{subfigure}[t]{0.15\textwidth}        
%		\centering
%		\begin{tikzpicture}
%			\setcounter{index}{0}
%			\coordinate (s) at (0,0);
%			\foreach \num in {$2$, $\infty$, $\infty$}{
%				\node[minimum size=6mm, draw, rectangle] at (s) {\num};
%				\node at ($(s)-(0,0.5)$) {\theindex};
%				\stepcounter{index}
%				\coordinate (s) at ($(s) + (0.6,0)$);
%			}
%		\end{tikzpicture}
%		%\caption{Leaf nodes of $\segtree_{1}^{4}$.}
%		\caption{Array $\dsarray{2}{3}$.}
%		\label{fig:interface-out2}
%	\end{subfigure}
%	~~
%	\begin{subfigure}[t]{0.15\textwidth}        
%		\centering
%		\begin{tikzpicture}
%			\setcounter{index}{0}
%			\coordinate (s) at (0,0);
%			\foreach \num in {$\infty$, \textcolor{bred}{$2$}, $\infty$}{
%				\node[minimum size=6mm, draw, rectangle] at (s) {\num};
%				\node at ($(s)-(0,0.5)$) {\theindex};
%				\stepcounter{index}
%				\coordinate (s) at ($(s) + (0.6,0)$);
%			}
%		\end{tikzpicture}
%		%\caption{Leaf nodes of $\segtree_{1}^{4}$.}
%		\caption{Array $\dsarray{0}{3}$.}
%		\label{fig:interface-out3}
%	\end{subfigure}

	\caption{Query $\suc(\graphnode{0}{0}, 3)$.}
	\label{fig:ds-full-example}
\end{figure}

\begin{example}
Consider the chain DAG in \cref{fig:ds-full-example} and the operation $\suc(\graphnode{0}{0}, 3)$. 
The first step is identifying the earliest successors of $\graphnode{0}{0}$ reachable via direct edges from $\graphnode{0}{0}$ and its thread-local successors, in all other threads.
This discovers $\graphnode{1}{0}$ via \textnormal{$\color{bred}{\circled{1}}$} and $\graphnode{3}{2}$ via \textnormal{$\color{bred}{\circled{2}}$}. 
While $\graphnode{3}{2}$ is indeed in chain $3$, it is not necessarily the \emph{earliest} successor of $\graphnode{0}{0}$ (indeed, the earliest successor is $\graphnode{3}{1}$).
Hence the function proceeds  iteratively for each newly discovered node until reaching a fixed point.
This leads to the discovery of $\graphnode{2}{1}$ via \textnormal{$\color{bred}{\circled{3}}$} and eventually the discovery of $\graphnode{3}{1}$ via \textnormal{$\color{bred}{\circled{4}}$}. 
%Ultimately, the node $\graphnode{3}{1}$ is returned.
\end{example}

%% file: sections/ds-incremental.tex
\section{Incremental $\ds$}\label{sec:incremental_ds}

In practice, many dynamic analyses maintain their partial order incrementally, i.e., they only insert and never delete orderings.
To optimize for this incremental setting, in this section we specialize $\ds$ to only handle incremental updates, with guarantees stated in the following theorem.

\begin{restatable}{theorem}{thmdsincremental}\label{thm:ds_incremental}
Consider a chain DAG $G$ of $\numEvents$ nodes, $\numThreads$ chains, and cross-chain density $\CrossChainDensity$.
Incremental $\ds$ maintain $G$ under incremental updates, costing $O(\numThreads^2 \min(\log \numEvents, \CrossChainDensity))$ time per update and $O(\min(\log \numEvents, \CrossChainDensity))$ time per query.
\end{restatable}

Compared to \cref{thm:ds_dynamic}, \cref{thm:ds_incremental} trades the cost dependency on the number of chains  from queries to updates, while also decreasing it by a factor $\numThreads$.
Moreover, now the cost of both queries and updates is bounded by the cross-chain density of $G$ (as opposed to only queries in \cref{thm:ds_dynamic}).

\myparagraph{Incremental $\ds$}
Incremental $\ds$ maintain $\numThreads(\numThreads -1)$ suffix minima arrays $\dsarray{t_1}{t_2}$, with $t_1\neq t_2$, each implemented using an $\sparsest$ $T_{t_1}^{t_2}$.
These arrays now store \emph{transitive reachability}, as opposed to direct edges between chains.
\cref{algo:ds-inc-interface} shows how each array $\dsarray{t_1}{t_2}$ is queried and updated, following the reachability queries and edge insertions in $\ds$.
The functions $\suc(\graphnode{t_1}{ j_1}, t_2)$,  $\pred(\graphnode {t_1}{ j_1} \rangle, t_2)$ and $\reachable(\graphnode{t_1}{ j_1}, \graphnode{t_2}{ j_2})$ handle successor, predecessor, and reachability queries in a straightforward manner:~since each $\dsarray{t_1}{t_2}$ is transitively closed, these operations reduce to suffix-minima queries on $\dsarray{t_1}{t_2}$.

The function $\insedge(\graphnode{t_1}{ j_1} , \graphnode{t_2}{ j_2 })$ is somewhat more complex, to guarantee that the arrays $\dsarray{t_1}{t_2}$ indeed store transitive reachability across chains.
To achieve this, besides inserting the direct edge $\graphnode{t_1}{ j_1} \to \graphnode{t_2}{ j_2 }$, this function performs a transitive closure computation by
finding the predecessor $\graphnode{t'_1}{j'_1}$ of $\graphnode{t_1}{j_1}$ (\crefrange{line:ds-inc-interface-ins-3}{line:ds-inc-interface-ins-4}) and the successor $\graphnode{t'_2}{j'_2}$ of  $\graphnode{t_2}{j_2}$ (\crefrange{line:ds-inc-interface-ins-5}{line:ds-inc-interface-ins-6}) in all pairs of chains $t'_1$, $t'_2$.
Then, an edge $\graphnode{t'_1}{j'_1} \to \graphnode{t'_2}{j'_2}$ is inserted unless a path between these nodes already exists.

\input{material/algorithms/opt-ds/interface-incremental}

\input{material/examples/ds-inc-interface}

\myparagraph{Analysis} 
The correctness of incremental $\ds$ is based on the following invariants.
The first concerns soundness, i.e., every entry in each array $\dsarray{t_1}{t_2}$ corresponds to a path between the respective nodes in chains $t_1$ and $t_2$.
The second concerns completeness, i.e., if there is a path $\graphnode{t_1}{j_1} \reach \graphnode{t_2}{j_2}$ between two nodes,
it is captured as a suffix minima in the array $\dsarray{t_1}{t_2}$.
In conjunction, they imply the correctness of $\FunctionSucc$, $\FunctionPred$, and $\FunctionQuery$ queries implemented as $\mn$ and $\argleq$ operations on the corresponding array $\dsarray{t_1}{t_2}$.

\begin{restatable}[Soundness]{lemma}{lemdsincsoundness}\label{lem:ds-inc-soundness}
$\dsarray{t_1}{t_2}[j_1] = j_2$ $\implies$ $\graphnode{t_1}{j_1} \reach \graphnode{t_2}{j_2}$.
\end{restatable}

\begin{restatable}[Completeness]{lemma}{lemdsinccompleteness}\label{lem:ds-inc-completeness}
$\graphnode{t_1}{j_1} \reach \graphnode{t_2}{j_2}$ and $t_1 \neq t_2$ $\implies$ $\exists j_{1}' \geq j_1$. $\dsarray{t_1}{t_2}[j_{1}'] \leq j_2$.
\end{restatable}

Recall that the key feature of $\sparsest$s is their ability to handle sparse arrays efficiently.
As incremental $\ds$ insert transitive edges in the arrays $\dsarray{t_1}{t_2}$, this might increase their density, adversely affecting the underlying $\sparsest$, as we saw in \cref{subsec:sparse_segment_tree}.
At closer inspection, however, the density of each $\dsarray{t_1}{t_2}$ cannot grow beyond the cross-chain density of $G$.

\begin{restatable}[Sparsity]{lemma}{lemdsincsparsity}\label{lem:ds-inc-sparsity}
Let $d$ be the cross-chain density of $G$.
Then the density of each array $\dsarray{t_1}{t_2}$ is bounded by $d$.
\end{restatable}

Combined with our bounds on the height of $\sparsest$s (\cref{lem:sparse_tree_height} in \cref{subsec:sparse_segment_tree}), \cref{lem:ds-inc-sparsity} implies that the height of the $\sparsest$ representing $\dsarray{t_1}{t_2}$ is bounded both by $\log \numEvents$ and by the cross-chain density of $G$. 
We thus arrive at \cref{thm:ds_incremental}.

\input{material/figures/bar_plots}

\myparagraph{Space usage}
Due to \cref{lem:ds-inc-sparsity}, the total space used by the suffix minima arrays $\dsarray{t_1}{t_2}$ is $O(\CrossChainDensity \numThreads)$, by an analysis similar to that of fully dynamic $\ds$ in \cref{subsec:ds-cssts}.
This time, $\ds$ does not use heaps to store all edges, thus the total space used is $O(\CrossChainDensity \numThreads)$.
Vector Clocks and $\naiveds$~\cite{Pavlogiannis-2019}, on the other hand, use $O(\numEvents\numThreads)$ space, which is larger in practice.

%% file: material/algorithms/opt-ds/interface-incremental.tex
%!TEX root = ../../../main.tex
%\setlength{\textfloatsep}{10pt}
\begin{algorithm}[t]
	\small
	\DontPrintSemicolon
	\SetInd{0.3em}{0.3em}
	
	\BlankLine
	\MyFunction{\FunctionSucc{$\graphnode{t_1}{ j_1}, t_2$}}{\label{line:ds-inc-interface-succ-1}
		\Return $\mn(\dsarray{t_1}{t_2}, j_1)$ \label{line:ds-inc-interface-succ-2}
	}
	
	\BlankLine
	\BlankLine
	\MyFunction{\FunctionPred{$\graphnode{t_1}{ j_1}, t_2$}}{\label{line:ds-inc-interface-pred-1}
		\Return $\argleq(\dsarray{t_2}{t_1}, j_1)$
		\label{line:ds-inc-interface-pred-2}
		%\footnote{This has a typo in \cite{popl19-pavlogiannis}.}
	}
	
	\BlankLine
	\BlankLine
	\MyFunction{\FunctionQuery{$\graphnode{t_1}{ j_1}, \graphnode{t_2}{ j_2}$}}{\label{line:ds-inc-interface-query-1}
		\lIfElse{$\suc(\graphnode{ t_1}{ j_1}, t_2) \leq j_2$}{\Return True}{\Return False}\label{line:ds-inc-interface-query-2}
	}
	
	\BlankLine
	\BlankLine
	\MyFunction{\FunctionInsertEdge{$\graphnode{t_1}{j_1}, \graphnode{t_2}{j_2}$}}{
		\ForEach{$t_{1}' \in [k]$}{\label{line:ds-inc-interface-ins-1}
			\ForEach{$t_{2}' \in [k] \setminus \set{t_1'}$}{\label{line:ds-inc-interface-ins-2}
				
				\lIf{$t_{1}' = t_1$}{Let $j_{1}' \xleftarrow[]{} \graphnode{t_1}{ j_1}$\label{line:ds-inc-interface-ins-3}}
				\lElse{Let $j_{1}' \xleftarrow[]{} \pred(\graphnode{t_1}{ j_1}, t_{1}')$\label{line:ds-inc-interface-ins-4}} 
				
				\lIf{$t_{2}' = t_2$}{Let $j_{2}' \xleftarrow[]{} \graphnode{t_2}{j_2}$\label{line:ds-inc-interface-ins-5}}
				\lElse{Let $j_{2}' \xleftarrow[]{} \suc(\graphnode{t_2}{j_2}, t_{2}')$\label{line:ds-inc-interface-ins-6}}

				\lIf{$\suc(\graphnode{t_{1}'}{j_{1}'}, t_{2}') > j_{2}'$} {\label{line:ds-inc-interface-ins-7}
					$\upd(\dsarray{t_{1}'}{t_{2}'}, j_{1}', j_{2}')$\label{line:ds-inc-interface-ins-8}
				}
			}
		}
	}
	
	\caption{
		Incremental $\ds$. 
	}
	\label{algo:ds-inc-interface}
\end{algorithm}
\normalsize

%% file: material/examples/ds-inc-interface.tex
%!TEX root = ../../main.tex

\begin{figure}%[h!]
	\centering
	\def\ystep{0.8}
	\begin{subfigure}[t]{0.45\textwidth}
		\centering
		\scalebox{0.95}{
			\definecolor{offwhite}{HTML}{000000}
			\tikzset{
				> = stealth,
				every node/.append style = {
					text = offwhite
				},
				every path/.append style = {
					arrows = ->,
					draw = offwhite
				},
			}
			\tikz{
				\node (a) at (0,-0.3) {Thread $0$};
				\node (a0) at (0,-1*\ystep) {$\graphnode{0}{0}$};
				\node (a1) at (0,-2*\ystep) {$\graphnode{0}{1}$};
				\node (a2) at (0,-3*\ystep) {$\graphnode{0}{2}$};
				\path (a0) edge (a1);
				\path (a1) edge (a2);
				
				\node (b) at (2,-0.3) {Thread $1$};
				\node (b0) at (2,-1*\ystep) {$\graphnode{1}{0}$};
				\node (b1) at (2,-2*\ystep) {$\graphnode{1}{1}$};
				\node (b2) at (2,-3*\ystep) {$\graphnode{1}{2}$};
				\path (b0) edge (b1);
				\path (b1) edge (b2);
				
				\node (c) at (4,-0.3) {Thread $2$};
				\node (c0) at (4,-1*\ystep) {$\graphnode{2}{0}$};
				\node (c1) at (4,-2*\ystep) {$\graphnode{2}{1}$};
				\node (c2) at (4,-3*\ystep) {$\graphnode{2}{2}$};
				\path (c0) edge (c1);
				\path (c1) edge (c2);
				
				\node (d) at (6,-0.3) {Thread $3$};
				\node (d0) at (6,-1*\ystep) {$\graphnode{3}{0}$};
				\node (d1) at (6,-2*\ystep) {$\graphnode{3}{1}$};
				\node (d2) at (6,-3*\ystep) {$\graphnode{3}{2}$};
				\path (d0) edge (d1);
				\path (d1) edge (d2);
				
				\path (a1) edge (b0);
				\path (b2) edge (c1);
				\path (c0) edge (d2);
				
				\draw[mosolid] (b1) to (c0);
				\draw[mo, bend right=30] (a1) to (d2);
			}
		} 
		\caption{A chain DAG $G$.}
		\label{fig:interface-dag}
	\end{subfigure}
	\\
	\begin{subfigure}[t]{0.14\textwidth}        
		\centering
		\begin{tikzpicture}
			\setcounter{index}{0}
			\coordinate (s) at (0,0);
			\foreach \num in {$\infty$, $0$, $\infty$}{
				\node[minimum size=6mm, draw, rectangle] at (s) {\num};
				\node at ($(s)-(0,0.5)$) {\theindex};
				\stepcounter{index}
				\coordinate (s) at ($(s) + (0.6,0)$);
			}
		\end{tikzpicture}
		\caption{Array $\dsarray{0}{1}$.}
		\label{fig:interface-out1}
	\end{subfigure}
	~~
	\begin{subfigure}[t]{0.15\textwidth}        
		\centering
		\begin{tikzpicture}
			\setcounter{index}{0}
			\coordinate (s) at (0,0);
			\foreach \num in {$2$, $\infty$, $\infty$}{
				\node[minimum size=6mm, draw, rectangle] at (s) {\num};
				\node at ($(s)-(0,0.5)$) {\theindex};
				\stepcounter{index}
				\coordinate (s) at ($(s) + (0.6,0)$);
			}
		\end{tikzpicture}
		\caption{Array $\dsarray{2}{3}$.}
		\label{fig:interface-out2}
	\end{subfigure}
	~~
	\begin{subfigure}[t]{0.15\textwidth}        
		\centering
		\begin{tikzpicture}
			\setcounter{index}{0}
			\coordinate (s) at (0,0);
			\foreach \num in {$\infty$, \textcolor{bred}{$2$}, $\infty$}{
				\node[minimum size=6mm, draw, rectangle] at (s) {\num};
				\node at ($(s)-(0,0.5)$) {\theindex};
				\stepcounter{index}
				\coordinate (s) at ($(s) + (0.6,0)$);
			}
		\end{tikzpicture}
		%\caption{Leaf nodes of $\segtree_{1}^{4}$.}
		\caption{Array $\dsarray{0}{3}$.}
		\label{fig:interface-out3}
	\end{subfigure}
	\caption{Inserting an edge $\graphnode{1}{1}\to \graphnode{2}{0}$ and inferring the transitive path $\graphnode{0}{1} \reach \graphnode{3}{2}$.}
	\label{fig:interface}
\end{figure}

\begin{example}
	Consider the chain DAG in \cref{fig:interface-dag} and the operation $\insedge(\graphnode{1}{1}, \graphnode{2}{0})$.
	We obtain the transitive path $\graphnode{0}{1} \reach \graphnode{3}{2}$ as follows.
	First, the predecessor of $\graphnode{1}{1}$ in thread $0$ is identified via the query $\pred(\graphnode{1}{1}, 0)$.
	This query performs an $\argleq$ operation on $\dsarray{0}{1}$ (\cref{fig:interface-out1}) with value $1$, returning node $\graphnode{0}{1}$.
	Then, the successor of $\graphnode{2}{0}$ in thread $3$ is identified via the query $\suc(\graphnode{2}{0}, 3)$.
	This query performs a $\mn$ operation on $\dsarray{2}{3}$ (\cref{fig:interface-out2}) with index $0$, and returns the node $\graphnode{3}{2}$.
	Next, the update operation on $\dsarray{0}{3}$ adds the transitive edge, resulting in $\dsarray{0}{3}$ in \cref{fig:interface-out3}. 
\end{example}

%% file: material/figures/bar_plots.tex
\begin{figure*}[h!]
\def\height{3.4cm}
\begin{subfigure}{0.68\textwidth}
\begin{tikzpicture}
\begin{axis}[
name=ax1,
inner sep=2pt,
ybar=2pt,
width=13cm,
height=\height,
ymin=0,
xtick=data,
xticklabels={Data Races, Deadlocks, Memory bugs, X86-TSO consistency, Use-after-free, C11 data races},
xticklabel style={rotate=10, font=\small}, % Rotate the x-axis labels by 45 degrees
ylabel={Mean Resource Ratio},
ylabel near ticks,
%ymode=log,
xlabel={},
bar width=10pt,
legend style={
at={(0.5,-0.55)},
anchor=north,legend columns=-1
},
ymajorgrids=true,
clip=false,
axis lines*=left,
ymax=2.2,
restrict y to domain*=0:2.6,
visualization depends on=rawy\as\rawy, % Save the unclipped values
after end axis/.code={ % Draw line indicating break
\draw [ultra thick, white, decoration={snake, amplitude=1pt}, decorate] (rel axis cs:0,1.05) -- (rel axis cs:1,1.05);
},
nodes near coords={\small \pgfmathprintnumber{\rawy}},
%nodes near coords,
% -----------------------------------------------------------------
% we create a style for the `nodes near coords` which is dependent
% on the value
% (adapted from <http://tex.stackexchange.com/a/141006/95441>)
% (#1: the THRESHOLD after which we switch to a special display)
nodes near coords greater equal only/.style={
% define the style of the nodes with "small" values
small value/.style={
/tikz/coordinate,
},
every node near coord/.append style={
check for small values/.code={
\begingroup
% this group is merely to switch to FPU locally.
% Might be unnecessary, but who knows.
\pgfkeys{/pgf/fpu}
\pgfmathparse{\pgfplotspointmeta<#1}
\global\let\result=\pgfmathresult
\endgroup
%
% simplifies debugging:
%\show\result
%
\pgfmathfloatcreate{1}{1.0}{0}
\let\ONE=\pgfmathresult
\ifx\result\ONE
% AH: our condition 'y < #1' is met.
\pgfkeysalso{/pgfplots/small value}
\fi
},
check for small values,
},
},
% assign a value to the new style which is the threshold at which
% the `small value` style is used.
nodes near coords greater equal only=2.6,
% -----------------------------------------------------------------
]

\addplot table [x expr=\coordindex, y=vct, col sep=comma,] {barplot1.csv};
\addplot table [x expr=\coordindex, y=stt, col sep=comma,] {barplot1.csv};
\addplot table [x expr=\coordindex, y=vcm, col sep=comma,] {barplot1.csv};
\addplot table [x expr=\coordindex, y=stm, col sep=comma] {barplot1.csv};

\legend{\small $\vc$ (time)~~, $\naiveds$ (time)\qquad\qquad, $\vc$ (memory)~~, $\naiveds$ (memory)}
\end{axis}
\end{tikzpicture}
\end{subfigure}
\begin{subfigure}{0.3\textwidth}
\begin{tikzpicture}
\begin{axis}[
inner sep=2pt,
ybar=2pt,
width=3cm,
height=\height,
ymin=0,
xtick=data,
xticklabels={Linearizability},
xticklabel style={rotate=10, font=\small}, % Rotate the x-axis labels by 45 degrees
ylabel={Mean Resource Ratio},
ylabel near ticks,
%ymode=log,
xlabel={},
bar width=10pt,
legend style={
at={(0.5,-0.55)},
anchor=north,legend columns=-1
},
ymajorgrids=true,
clip=false,
axis lines*=left,
ymax=13,
]

\addplot table [x expr=\coordindex, y=grapht, col sep=comma,] {barplot2.csv};
\addplot table [x expr=\coordindex, y=graphm, col sep=comma,] {barplot2.csv};

\legend{\small $\graphds$ (time)~~, $\graphds$ (memory)}
\end{axis}

\end{tikzpicture}
\end{subfigure}
\caption{
The geometric mean of the ratio of time and memory used in each benchmark by $\vc$ and $\naiveds$ over incremental $\ds$ (left), and $\graphds$ over fully dynamic $\ds$ (right), in the dataset of the corresponding analysis.
}
\label{fig:bars}
\end{figure*}

%% file: sections/experiments-main.tex
\input{sections/experiments-beginning}

\subsection{Experimental Results}\label{subsec:experimental_results}

We begin with a macroscopic view in \cref{fig:bars}, showing the (geometric) average improvement in time and memory that $\ds$ offer over $\vc$, $\naiveds$ and $\graphds$ in each analysis.
We see that in nearly all cases, these ratios are above 1, meaning that $\ds$ indeed outperform the other data structures.
The only exception concerns data races in C11; we will analyze this case in more detail later.
The improvement is even more pronounced on the last analysis (linearizability), which is fully dynamic and thus the only baseline is plain $\graphds$.

Although the worst-case memory usage of $\ds$ is $O(\numEvents\numThreads)$, and thus the same as $\vc$ and $\naiveds$, in practice it is less.
This confirms our insights that typical analyses give rise to sparse chain DAGs,
which is then exploited by $\ds$.
%One interesting point is the far higher memory usage of $\vc$ in the case of X86-TSO.
We will take a closer look into the sparsity of each benchmark later.

We remark that the resource (time, memory) reported in each benchmark concerns the whole analysis, not just the corresponding data structure\footnote{The analysis on data races for C11 is an exception, as we explain later.}.
The actual resource improvement achieved by $\ds$ is significantly larger.

\input{sections/experiments-race_prediction}
\input{sections/experiments-deadlock_prediction}
\input{sections/experiments-memory_bug_prediction}
\input{sections/experiments-consistency-checking}

\input{sections/experiments-ufo}
\input{sections/experiments-c11tester}
\input{sections/experiments-root-causing}

\input{material/figures/scalability}

\subsection{Scalability Analysis} \label{subsec:scalability_analysis}

Here we perform controlled scalability experiments on the performance of $\ds$ for edge insertions and queries.
We consider partial orders $P$ in two settings, consisting of $\numThreads=10,20$ chains, and varying number of events $\numChainEvents$ in each chain.
Initially, $P$ has no cross-chain edges.
First, to measure the performance of edge insertions,
we repeatedly insert random cross-chain edges $\graphnode{t}{i} \mapsto \graphnode{t'}{j}$ such that  the two endpoints are unordered and $|i-j| \leq b$.
The last condition captures the fact that cross-chain orderings are typically between events that execute within the same time-window.
We report on window $b=10^4$, though we have observed similar results with other values of $b$.
This condition also prevents $P$ from becoming close to total.
We attempt to insert $20 \numChainEvents$ edges in each case,
and report the average time of edge insertions.
Second, to measure the performance of edge queries, we use the partial order $P$ at the end of the above process.
We make $10^6$ random queries, and report the average time per query.

The results are shown in \cref{fig:scalability}.
As expected by theory, $\vc$ suffer linear complexity for insertions but achieve queries in constant time.
On the other hand, $\ds$ and $\naiveds$ achieve logarithmic complexity in both insertions and deletions.
$\ds$ are far more performant than $\vc$ in edge insertions, but also achieve query times that are similar to $\vc$, though, naturally, $\vc$ are slightly faster, as each query is a simple lookup.
This is achieved by our minima indexing and sparse representation, which makes queries run with minimal overhead in practice.
Finally, $\ds$ are considerably faster than $\naiveds$ in both insertions and queries, an improvement stemming from the Sparse Segment Tree (\cref{subsec:sparse_segment_tree}).

%% file: sections/experiments-beginning.tex
\section{Experimental Evaluation}\label{sec:experiments}
Here we incorporate $\ds$ in a number of dynamic analyses from recent literature and evaluate their performance.

\subsection{Experimental Setup}\label{subsec:experimental_setup}

\myparagraph{Data structures}
Each analysis dynamically maintains a partial order $P$ using various representations.
We explore different standard ways for this purpose, as follows.
\begin{compactitem}
\item $\graphds$ is a standard graph representation of $P$, that is not transitively closed.
\item $\vc$ is a standard Vector Clock representation of $P$, that is (by design) transitively closed~\cite{Mattern-1989}.
\item $\naiveds$ is the data structure in~\cite{Pavlogiannis-2019} based on Segment Trees.
\item $\ds$, as introduced in this paper.
\end{compactitem}
Most analyses make only incremental updates, in which case Vector Clocks are preferred over plain graphs.
Thus in this setting we compare $\ds$ to $\vc$ and $\segtree$.
On the other hand, $\vc$ and $\segtree$ cannot handle decremental updates, hence in the fully dynamic setting we compare $\ds$ to $\graphds$.
The data structure used in the respective paper is marked by $\dagger$.
As some tools are not public, we implemented their analysis following the corresponding paper.

\myparagraph{Optimizations}
To set the threshold $\threshold$ for the size of the block nodes in $\ds$ (\cref{subsec:sparse_segment_tree}), we perform a randomized stress test with varying sizes of $\threshold$, and measure their performance.
%Although this is not exhaustive, it gives as some indication on how to optimize this value.
%The results are shown in \cref{subsec:app_randomized_queries}.
Based on this test, we set $\threshold=32$.

We also implement two optimizations in $\vc$.
First, each new edge $e_1\to e_2$ must be propagated to the successors of $e_2$ in its own chain.
We stop this propagation as soon as we find some $e'_2$ with already $e_1\reach e'_2$.
Second, for each chain, we do not create $\vc$ for the suffix of its events that do not have any direct orderings from other chains, as ordering queries on such nodes can be inferred from their predecessors.

\myparagraph{Setup}
In each analysis, we use the dataset of the corresponding paper and relevant literature,
%\footnote{Some benchmarks were inaccessible or produced compilation errors.}, 
and report on benchmarks on which some data structure runs in $\geq 1$s.
%For completeness, all benchmarks are reported in \cref{subsec:app_full_tables}.
We use an Ubuntu 22.04 machine with 2.4GHz CPU and 64GB of memory.

%% file: sections/experiments-race_prediction.tex
%\subsection{Data Race Prediction}\label{subsec:data_race_prediction}
\myparagraph{1.~Data race prediction}
We start with the $\mtwo$ data race detector~\cite{Pavlogiannis-2019}.
This analysis observes traces $\tr$ which might be data-race free, and attempts to permute them to new traces $\tr'$ that are valid (called correct reorderings) and expose a data race.
Internally, $\mtwo$  utilizes $\naiveds$ for maintaining its partial order $P$.
The reachability queries are incremental.
We measure the total time taken in the analysis.

\cref{tbl:m2} shows the performance of $\mtwo$ with each data structure.
$\vc$ become unscalable early, and are outperformed by $\naiveds$, which were indeed developed to address this scalability issue.
We observe that $\ds$ are the most performant data structure on each individual benchmark.
Their advantage over $\naiveds$ is primarily due to the efficient way that $\ds$ handle sparsity (\cref{subsec:sparse_segment_tree}).
Indeed, column $\AvgDensity$ reports the mean density among each suffix minima array inside $\ds$ when it obtained its densest form, and matches our expectations that is typically small.
Our sparse treatment not only reduces the runtime of $\ds$ but also their memory footprint, and thus support the analysis even when $\vc$ and $\naiveds$ go out of memory (on \texttt{xalan} and \texttt{batik}).

As a side note, observe that the running time of each method is not always increasing when we move to larger traces
(eg, all data structures take more time to process \texttt{moldyn}, with size $N=200.3$K than \texttt{jigsaw}, with size $3.1$M).
This is because \texttt{moldyn} has more potential data races that the analysis needs to check, despite its smaller size.
Similar, non-monotonic, patterns appear in other analyses as well.

\input{material/tables/m2-result.tex}

%% file: material/tables/m2-result.tex
%!TEX root = ../../main.tex

\begin{table}[h!]
    \caption{
    %Race prediction results. Columns 1-2 denote the number of threads and events, respectively.
    %Columns 3-5 show the times taken in seconds. Timeout is set to 5h.
    Race prediction results. The timeout is 5h.
    }
    %\vspace{-0.15cm}
    \setlength\tabcolsep{3pt}
    \small
    \centering
    \scalebox{1}{
    \begin{tabular}{|r|c|c|c|c|c|c|}
\hline
\textbf{benchmark} & $\mathit{T}$ & $\mathit{N}$ & $\AvgDensity$ & $\vc$ (s) & $\naiveds^\dagger$ (s) & $\ds$ (s)\\
\hline\hline
clean & 12 & 1.3K & 0.28 & 2.4 & 2.6 & 1.6\\
bubblesort & 29 & 4.7K & 0.15 & 69.4 & 56.2 & 27.5\\
lang & 10 & 6.3K & 0.22 & 93.3 & 39.2 & 27.3\\
readerswriters & 8 & 11.3K & 0.32 & 54.4 & 23.5 & 18.5\\
raytracer & 6 & 15.8K & 0.17 & 10.5 & 10.2 & 9.3\\
bufwriter & 9 & 22.3K & 0.2 & 130 & 24.6 & 12.9\\
ftpserver & 14 & 49.6K & 0.06 & 41.2 & 14.1 & 9.0\\
moldyn & 6 & 200.3K & 0.12 & T.O. & T.O. & 12636\\
linkedlist & 15 & 1.0M & 0.06 & T.O. & T.O. & 8893\\
derby & 7 & 1.4M & 0.04 & 1436 & 215 & 197\\
jigsaw & 15 & 3.1M & 0.06 & 154 & 45.6 & 32.0\\
sunflow & 17 & 11.7M & 0.01 & T.O. & 780 & 505\\
xalan & 9 & 122.5M & 0.01 & T.O. & O.O.M. & 979\\
batik & 8 & 157.9M & 0.01 & O.O.M. & O.O.M. & 1956\\
\hline\hline\textbf{Total} & \textbf{-} & \textbf{297.9M} & \textbf{-} & \textbf{>91992} & \textbf{>73211} & \textbf{25304}\\
                    \hline
                \end{tabular}
            }
        \label{tbl:m2}
        \end{table}
        

%% file: sections/experiments-deadlock_prediction.tex
\myparagraph{2.~Deadlock prediction}
Here we incorporate $\ds$ in $\seqcheck$, a dynamic deadlock prediction analysis~\cite{Cai-2021}.
%\footnote{As the tool is not public, we have implemented the analysis following~\cite{Cai-2021}.}.
This analysis identifies potential deadlocks by analyzing lock-acquisition orders between threads, and then try to witness each deadlock by a valid reordering of the observed trace. 
For this purpose, it incrementaly maintains a partial order which will eventually be linearized to this valid reordering.

Our results are shown in \cref{tbl:dl}. 
$\vc$ are again the slowest performer by far, though they are very competitive on a few benchmarks for which deadlocks can be detected fairly early in the analysis.
Naturally, $\naiveds$ and $\ds$ perform similarly on these benchmarks as well.
On the more demanding benchmarks, however, $\ds$ is clearly the best performer.
We also observe that the average density $\AvgDensity$ is small.

\input{material/tables/deadlock-result.tex}

%% file: material/tables/deadlock-result.tex
%!TEX root = ../../main.tex

\begin{table}[h!]
    \caption{
    %Consistency checking results.  Columns 1-2 denote the number of threads and events, respectively.
    %Columns 3-5 show the times taken in seconds. Timeout is set to 1h.
    Deadlock results. %The timeout is 1h.
    }
    %\vspace{-0.15cm}
    \setlength\tabcolsep{3pt}
    \small
    \centering
    \scalebox{1}{
    \begin{tabular}{|r|c|c|c|c|c|c|}
\hline
%\textbf{benchmark} & $\mathit{T}$ & $\mathit{N}$ & $\AvgDensity$ & VCs & STs & CSSTs\\
%\hline\hline
\textbf{benchmark} & $\mathit{T}$ & $\mathit{N}$ & $\AvgDensity$ & $\vc$ (s) & $\naiveds^\dagger$ (s) & $\ds$ (s)\\
\hline\hline
jigsaw & 21 & 143.0K & 0.03 & 356 & 14.8 & 7.8\\
elevator & 5 & 245.9K & 0.02 & 3.7 & 2.8 & 2.6\\
hedc & 7 & 409.8K & 0.04 & 23.2 & 22.0 & 23.2\\
JDBCMySQL & 3 & 442.9K & 0.01 & 3.7 & 4.1 & 3.5\\
cache4j & 2 & 775.5K & 0.01 & 5.5 & 6.6 & 5.6\\
Swing & 8 & 3.8M & 0.01 & 14.8 & 13.4 & 13.1\\
sunflow & 15 & 21.5M & 0.01 & 369 & 111 & 110\\
eclipse & 15 & 64.2M & 0.01 & 535 & 711 & 282\\
\hline\hline\textbf{Total} & \textbf{-} & \textbf{91.4M} & \textbf{-} & \textbf{1311} & \textbf{886} & \textbf{448}\\
        \hline
        \end{tabular}
        }
    \label{tbl:dl}
    \end{table}
        

%% file: sections/experiments-memory_bug_prediction.tex
%\subsection{Memory Bug Prediction}
\myparagraph{3.~Memory bug prediction}
Here we incorporate $\ds$ in $\convulpoe$, a dynamic analysis targeting memory bugs due to concurrency~\cite{Yu-2021}, by reordering observed traces.
%\footnote{As the tool is not public, we have implemented the analysis following~\cite{Yu-2021}.}.
%$\convulpoe$ observes traces and attempts to reorder them so as to expose a bug.
%Internally, it utilizes $\naiveds$ for handling reachability.
The reachability queries are incremental.
We measured the total time taken in the analysis.

Our results are shown in \cref{tbl:convulpoe}.
We again observe that $\vc$ are the slowest data structure to support this analysis.
While $\naiveds$ offer a generous speedup, $\ds$ are again the most performant, and even manage to handle benchmarks that makes $\vc$ to time out and $\naiveds$ to go out of memory.

\input{material/tables/convulpoe-result.tex}

%% file: material/tables/convulpoe-result.tex
%!TEX root = ../../main.tex

\begin{table}[h!]
    \caption{
    %Memory bug prediction results. Columns 1-2 denote the number of threads and events, respectively.
    %Columns 3-5 show the times taken in seconds. Timeout is set to 5h.
    Memory bug prediction results. The timeout is 5h.
    }
    %\vspace{-0.15cm}
    \setlength\tabcolsep{3pt}
    \small
    \centering
    \scalebox{1}{
    \begin{tabular}{|r|c|c|c|c|c|c|}
\hline
\textbf{benchmark} & $\mathit{T}$ & $\mathit{N}$ & $\AvgDensity$ & $\vc$ (s) & $\naiveds^\dagger$ (s) & $\ds$ (s)\\
\hline\hline
pbzip2 & 7 & 243.5K & 0.06 & 298 & 6.7 & 3.8\\
pigz & 6 & 2.3M & 0.01 & T.O. & 2087 & 1265\\
xz & 2 & 3.0M & 0.01 & 15.1 & 13.9 & 12.9\\
lbzip2 & 11 & 9.3M & 0.04 & 273 & 96.6 & 46.9\\
x264 & 7 & 10.4M & 0.03 & 474 & 165 & 62.0\\
libvpx & 2 & 19.0M & 0.01 & 440 & 83.6 & 77.6\\
libwebp & 2 & 29.5M & 0.1 & 324 & 118 & 115\\
x265 & 15 & 37.1M & 0.02 & T.O. & O.O.M. & 627\\
\hline\hline\textbf{Total} & \textbf{-} & \textbf{110.9M} & \textbf{-} & \textbf{>37824} & \textbf{>20570} & \textbf{2211}\\
    \hline
    \end{tabular}
    }
\label{tbl:convulpoe}
\end{table}
    

%% file: sections/experiments-consistency-checking.tex
\myparagraph{4.~Consistency checking in x86-TSO}
Here we report our results for consistency checking under the x86-TSO memory model.
Though the problem is NP-complete, polynomial-time heuristics are known to be remarkably accurate.
We experiment with the consistency analysis developed in~\cite{Roy-2006}.
%\footnote{As the tool is not public, we have implemented the analysis following~\cite{Roy-2006}.}.
In contrast to the previous analyses where the underlying chain DAG had one chain per thread, here we have two chains per thread, one for its program order and one for its store buffer.
The reachability queries are incremental.
We measure the total time taken in the analysis.

The results are shown in \cref{tbl:cc}.
$\vc$ are very slow compared to the other two data structures.
Indeed, this analysis is quite demanding, as it performs repeated updates between events that are in the middle of the partial order, which leads to deep edge propagations.
Although $\naiveds$ perform considerably better than $\vc$, $\ds$ are decisively faster in this setting too.
Finally, the average density $\AvgDensity$ is larger here than in previous analyses, but still considerably below $1$.

\input{material/tables/consistency-checking-result.tex}

%% file: material/tables/consistency-checking-result.tex
%!TEX root = ../../main.tex

\begin{table}[h!]
    \caption{
    %Consistency checking results.  Columns 1-2 denote the number of threads and events, respectively.
    %Columns 3-5 show the times taken in seconds. Timeout is set to 1h.
    Consistency checking results. The timeout is 1h.
    }
    %\vspace{-0.15cm}
    \setlength\tabcolsep{3pt}
    \small
    \centering
    \scalebox{1}{
    \begin{tabular}{|r|c|c|c|c|c|c|}
\hline
\textbf{benchmark} & $\mathit{T}$ & $\mathit{N}$ & $\AvgDensity$ & $\vc$ (s) & $\naiveds$ (s) & $\ds$ (s)\\
\hline\hline
dekker & 3 & 16.3K & 0.41 & 6.9 & 0.3 & 0.25\\
peterson & 3 & 19.0K & 0.37 & 9.4 & 0.4 & 0.45\\
lamport & 3 & 28.1K & 0.39 & 18.3 & 0.6 & 0.6\\
dq & 4 & 31.5K & 0.24 & 11.5 & 0.5 & 0.44\\
chase-lev & 5 & 32.2K & 0.24 & 23.3 & 0.6 & 0.51\\
szymanski & 3 & 77.9K & 0.33 & 17.4 & 0.9 & 0.8\\
buf-ring & 9 & 115.3K & 0.39 & 322 & 7.3 & 6.27\\
mcs-lock & 11 & 196.4K & 0.17 & 298 & 28.3 & 20.71\\
spsc & 3 & 243.6K & 0.53 & 2548 & 2.1 & 1.74\\
linuxrwlocks & 6 & 276.4K & 0.28 & T.O. & 12.6 & 11.18\\
fib-bench & 3 & 300.0K & 0.4 & T.O. & 4.7 & 4.21\\
seqlock & 17 & 318.1K & 0.15 & 1227 & 46.1 & 28.16\\
spinlock & 11 & 482.5K & 0.39 & T.O. & 103 & 75.11\\
ttaslock & 11 & 491.1K & 0.41 & T.O. & 111 & 81.19\\
exp-bug & 4 & 498.5K & 0.38 & 2591 & 3.6 & 2.76\\
mutex & 11 & 519.7K & 0.56 & T.O. & 122 & 86.09\\
ticketlock & 6 & 569.6K & 0.4 & T.O. & 20.8 & 17.17\\
gcd & 3 & 750.1K & 0.28 & T.O. & 6.0 & 4.52\\
indexer & 17 & 800.0K & 0.51 & 7.8 & 5.8 & 3.23\\
twalock & 11 & 900.0K & 0.43 & T.O. & 174 & 118.39\\
treiber & 6 & 1.0M & 0.22 & T.O. & 20.7 & 16.73\\
mpmc & 10 & 2.0M & 0.17 & T.O. & 41.6 & 22.59\\
barrier & 5 & 2.8M & 0.6 & T.O. & 139 & 112.61\\
\hline\hline\textbf{Total} & \textbf{-} & \textbf{12.5M} & \textbf{-} & \textbf{>46680} & \textbf{852} & \textbf{616}\\
        \hline
        \end{tabular}
        }
    \label{tbl:cc}
    \end{table}

%% file: sections/experiments-ufo.tex
\myparagraph{5.~Use-after-free prediction}
Here we incorporate $\ds$ in the $\ufo$ analysis for use-after-free vulnerabilities due to concurrency~\cite{Huang-2018}.
This analysis is based on SMT, but relies on efficient partial-order reasoning to generate SMT queries.
The reachability queries are incremental.
We measure the time to generate the queries.
Our results are shown in \cref{tbl:ufo}.

In alignment with our observations so far, $\ds$ outperform $\vc$ and $\naiveds$ on all benchmarks.
However the speedup is not as large as in other analyses.
Looking closer into the running times, we observe that the analysis itself spends considerable time in components other than maintaining reachability, which are common for all data structures. 
If we only focus on the time taken for maintaining the partial order, $\ds$ indeed offer a significant speedup, e.g., $3.5 \times$ and $1.7 \times$ in \texttt{BoundedBuffer} over $\vc$ and $\naiveds$, respectively.
 %spend $67$ seconds in reachability only, while $\vc$ and $\naiveds$ spend $236$ and $119$ seconds,

\input{material/tables/ufo-result.tex}

%% file: material/tables/ufo-result.tex
%!TEX root = ../../main.tex

\begin{table}[h!]
    \caption{
    %UFO results. Columns 1-2 denote the number of threads and events, respectively.
    %Columns 3-5 show the times taken in seconds. Timeout is set to 5h.
    Use-after-free results. The timeout is 5h.
    }
    %\vspace{-0.15cm}
    \setlength\tabcolsep{3pt}
    \small
    \centering
    \scalebox{1}{
    \begin{tabular}{|r|c|c|c|c|c|c|}
\hline
\textbf{benchmark} & $\mathit{T}$ & $\mathit{N}$ & $\AvgDensity$ & $\vc^\dagger$ (s) & $\naiveds$ (s) & $\ds$ (s)\\
\hline\hline
bbuf & 3 & 27.7K & 0.02 & 426 & 459 & 415\\
BoundedBuffer & 11 & 325.7K & 0.02 & 835 & 736 & 679\\
DiningPhil & 21 & 1.4M & 0.01 & 1559 & 1191 & 1110\\
fanger01-ok & 5 & 93.4K & 0.06 & 366 & 194 & 190\\
qtsort & 6 & 41.7M & 0.01 & 104 & 48.2 & 31.9\\
pbzip & 5 & 269.3K & 0.01 & T.O. & 15894 & 14471\\
\hline\hline\textbf{Total} & \textbf{-} & \textbf{43.8M} & \textbf{-} & \textbf{>21290} & \textbf{18521} & \textbf{16898}\\
    \hline
    \end{tabular}
    }
\label{tbl:ufo}
\end{table}
    

%% file: sections/experiments-c11tester.tex
\myparagraph{6.~Data-race detection in C11}
Here we incorporate $\ds$ in $\celeventester$, a data-race detector for the C11 memory model~\cite{Luo-2021}.
The analysis constructs incrementally a trace $\tr$, by iteratively mapping a read to a write to obtain its value from.
To make these choices consistent with the C11 memory model, it maintains a partial order $P$ using $\vc$.
The reachability queries are incremental.
%This is an on-the-fly analysis and each execution is non-deterministic.
To ensure that all data structures process the same trace, we run them alongside each other on a single execution, and measure the time taken to maintain the partial order in each.

Our results are shown in \cref{tbl:c11tester}.
Interestingly, $\vc$ are generally more performant than $\naiveds$ and $\ds$ in this setting.
Looking closer into each benchmark, we observe that new orderings lead to very little (typically none at all) propagation in $P$.
Effectively, this makes $\vc$ run in $O(1)$ time per update, and thus gain an advantage.
In contrast, in \texttt{readerswriters} and \texttt{atomicblocks} $\celeventester$ inserts non-trivial orderings in $P$ (i.e., orderings that lead to propagation),
and $\ds$ are considerably more performant than $\vc$.
Finally, the average density $\AvgDensity$ is high in a few benchmarks, reaching even the value $1$, which explains why $\ds$ and $\naiveds$ have comparable performance, though $\ds$ remain more performant.

\input{material/tables/c11tester-result.tex}

%% file: material/tables/c11tester-result.tex
%!TEX root = ../../main.tex

\begin{table}[h!]
    \caption{
    Race detection on C11.
    %$\celeventester$ results.  Columns 1-2 denote the number of threads and events, respectively.
    %Columns 3-5 show the times taken in seconds.
    %The benchmarks in which propagation is observed are marked with (*).
    }
    %\vspace{-0.15cm}
    \setlength\tabcolsep{3pt}
    \small
    \centering
    \scalebox{1}{
    \begin{tabular}{|r|c|c|c|c|c|c|}
\hline
\textbf{benchmark} & $\mathit{T}$ & $\mathit{N}$ & $\AvgDensity$ & $\vc^\dagger$ (s) & $\naiveds$ (s) & $\ds$ (s)\\
\hline\hline
dq & 5 & 72.8K & 0.48 & 2.5 & 2.8 & 3.1\\
mabain & 7 & 101.6K & 0.33 & 0.6 & 1.2 & 1.1\\
seqlock & 18 & 693.9K & 0.2 & 4.5 & 6.7 & 6.3\\
iris-1 & 13 & 1.1M & 0.2 & 54.8 & 66.5 & 60.3\\
qu & 11 & 1.3M & 0.43 & 5.9 & 7.5 & 7.2\\
indexer & 18 & 1.6M & 1 & 1.0 & 1.1 & 1.1\\
exp-bug & 5 & 2.5M & 0.25 & 2.6 & 2.5 & 2.5\\
twalock & 12 & 4.5M & 0.34 & 29.6 & 53.3 & 49.0\\
gcd & 4 & 4.5M & 1 & 2.6 & 2.8 & 2.7\\
spinlock & 12 & 4.8M & 0.2 & 23.0 & 37.9 & 34.0\\
ttaslock & 12 & 4.9M & 0.25 & 26.7 & 48.0 & 42.7\\
silo & 5 & 5.6M & 0.01 & 5.9 & 6.2 & 6.1\\
fib-bench & 4 & 6.0M & 1 & 6.5 & 6.3 & 6.5\\
linuxrwlocks & 7 & 6.9M & 0.39 & 30.8 & 40.6 & 37.7\\
barrier & 6 & 8.3M & 0.51 & 19.6 & 24.5 & 23.7\\
mpmc & 11 & 9.2M & 0.24 & 37.6 & 44.9 & 42.6\\
spsc & 4 & 9.7M & 0.55 & 12.2 & 12.2 & 12.1\\
mcs-lock & 12 & 9.9M & 0.37 & 55.2 & 86.8 & 80.1\\
treiber & 7 & 10.1M & 0.11 & 31.3 & 36.6 & 35.7\\
iris-2 & 4 & 11.5M & 0.2 & 15.5 & 14.7 & 14.7\\
gdax & 8 & 13.5M & 0.97 & 8.5 & 7.6 & 7.6\\
ticketlock & 7 & 14.2M & 0.48 & 45.1 & 70.4 & 64.2\\
mutex & 12 & 15.6M & 0.39 & 48.5 & 68.5 & 63.7\\
readerswriters & 13 & 10.1M & 0.33 & 16.3 & 7.6 & 7.5\\
atomicblocks & 33 & 15.5M & 0.5 & 9.7 & 1.7 & 1.6\\
\hline\hline\textbf{Total} & \textbf{-} & \textbf{172.4M} & \textbf{-} & \textbf{496} & \textbf{659} & \textbf{614}\\
            \hline
            \end{tabular}
            }
        \label{tbl:c11tester}
        \end{table}
            

%% file: sections/experiments-root-causing.tex
%\subsection{Root Causing}
\myparagraph{7.~Root-causing linearizability violations}
Here we incorporate $\ds$ in a recent root-cause analysis for linearizability violations~\cite{Cirisci-2020}.
The reachability queries are both incremental and decremental,
hence the only baseline for maintaining the partial order are plain $\graphds$, as indeed used in~\cite{Cirisci-2020}.
We measure the total time taken in the analysis.

Our results are shown in \cref{tbl:rc}.
The experimental setup of~\cite{Cirisci-2020} consists of three data structures, accessed an increasing number of times.
$\ds$ are typically orders of magnitude faster than plain $\graphds$ used in~\cite{Cirisci-2020},
which strongly supports $\ds$ as an efficient data structure also for analyses utilizing decremental partial-order updates.
\input{material/tables/rc-result.tex}

%% file: material/tables/rc-result.tex
%!TEX root = ../../main.tex

\begin{table}[h!]
    \caption{
    Root causing linearizability violation results.
    %Root causing results. Columns 1-2 denote the number of threads and events, respectively.
    %Columns 3-4 show the times taken in seconds. Timeout is set to 1h.
    }
    %\vspace{-0.15cm}
    \setlength\tabcolsep{3pt}
    \small
    \centering
    \scalebox{1}{
    \begin{tabular}{|r|c|c|c|c|c|}
\hline
\textbf{benchmark} & $\mathit{T}$ & $\mathit{N}$ & $\AvgDensity$ & $\graphds^{\dag}$ (s) & $\ds$ (s)\\
\hline\hline
\multirow{4}{*}{\shortstack[l]{LogicalOrderingAVL}} & 3 & 615 & 0.1 & 1.2 & 0.4\\
 & 3 & 1.6K & 0.07 & 6.1 & 1.6\\
 & 3 & 2.6K & 0.06 & 13.5 & 3.8\\
 & 3 & 5.0K & 0.05 & 49.5 & 10.8\\ \hline
 \multirow{4}{*}{\shortstack[l]{OptimisticListSorted-\\SetWaitFreeContains}} & 3 & 219 & 0.19 & 1.4 & 0.2\\
 & 3 & 399 & 0.18 & 9.8 & 0.4\\
 & 3 & 759 & 0.17 & 114 & 0.9\\
 & 3 & 1.1K & 0.17 & 512 & 1.7\\ \hline
\multirow{4}{*}{\shortstack[l]{RWLockCoarse-\\GrainedListIntSet}}  & 3 & 978 & 0.04 & 3.6 & 0.8\\  
& 3 & 1.6K & 0.03 & 11.4 & 1.7\\
 & 3 & 3.2K & 0.03 & 73.6 & 5.3\\
 & 3 & 4.8K & 0.03 & 249 & 10.1\\
\hline\hline\textbf{Total} & \textbf{-} & \textbf{22.9K} & \textbf{-} & \textbf{1045} & \textbf{38}\\
    \hline
    \end{tabular}
    }
\label{tbl:rc}
\end{table}

%% file: material/figures/scalability.tex
\begin{figure*}[!h]
\def\height{3.65cm}
\def\width{5.1cm}
\pgfplotsset{compat=1.15}
\scalebox{0.9}{%
\begin{tikzpicture}
\begin{groupplot}[
group style={group size= 4 by 2, vertical sep=1.4cm, horizontal sep=1.3cm} , 
every x tick scale label/.style={at={(1,0)},xshift=1pt,anchor=south west,inner sep=0pt},
legend columns=-1
]
\nextgroupplot[
inner sep=2pt,
width=\width,
height=\height,
ylabel={Time (s)},
ylabel near ticks,
xlabel={$\numChainEvents$},
grid=both,
clip=false,
axis lines*=left,
title={Insert, $\numThreads=10$},
ymin={0},
xtick distance=1e4,
ytick distance=1e-3,
]

\addplot[mark=*, mark size=2pt,color=black] table[x=x, y=cssts, col sep=comma]{scalability_vcs_insert_k=10_r=10000_i=20n.csv};\label{plots:curve_cssts}
\addplot[mark=square*, mark size=2pt,color=teal] table[x=x, y=vcs, col sep=comma]{scalability_vcs_insert_k=10_r=10000_i=20n.csv};\label{plots:curve_vcs}
\coordinate (top) at (rel axis cs:0,1);% coordinate at top of the first plot

\nextgroupplot[
inner sep=2pt,
width=\width,
height=\height,
ylabel={Time (s)},
ylabel near ticks,
xlabel={$\numChainEvents$},
grid=both,
clip=false,
axis lines*=left,
title={Insert, $\numThreads=20$},
ymin={0},
xtick distance=1e4,
ytick distance=5e-3,
]

\addplot[mark=*, mark size=2pt,color=black] table[x=x, y=cssts, col sep=comma]{scalability_vcs_insert_k=20_r=10000_i=20n.csv};
\addplot[mark=square*, mark size=2pt,color=teal] table[x=x, y=vcs, col sep=comma]{scalability_vcs_insert_k=20_r=10000_i=20n.csv};

\nextgroupplot[
inner sep=2pt,
width=\width,
height=\height,
ylabel={Time (s)},
ylabel near ticks,
xlabel={$\numChainEvents$},
grid=both,
clip=false,
axis lines*=left,
title={Query, $\numThreads=10$},
ymin={0},
xtick distance=1e4,
ytick distance=2.5e-7,
]

\addplot[mark=*, mark size=2pt,color=black] table[x=x, y=cssts, col sep=comma]{scalability_vcs_query_k=10_r=10000_i=20n.csv};
\addplot[mark=square*, mark size=2pt,color=teal] table[x=x, y=vcs, col sep=comma]{scalability_vcs_query_k=10_r=10000_i=20n.csv};

\nextgroupplot[
inner sep=2pt,
width=\width,
height=\height,
ylabel={Time (s)},
ylabel near ticks,
xlabel={$\numChainEvents$},
grid=both,
clip=false,
axis lines*=left,
title={Query, $\numThreads=20$},
ymin={0},
xtick distance=1e4,
ytick distance=2.5e-7,
]

\addplot[mark=*, mark size=2pt,color=black] table[x=x, y=cssts, col sep=comma]{scalability_vcs_query_k=20_r=10000_i=20n.csv};
\addplot[mark=square*, mark size=2pt,color=teal] table[x=x, y=vcs, col sep=comma]{scalability_vcs_query_k=20_r=10000_i=20n.csv};
\coordinate (c2) at (rel axis cs:1,1);

%%%%%%%%%% STs

\nextgroupplot[
inner sep=2pt,
width=\width,
height=\height,
ylabel={Time (s)},
ylabel near ticks,
xlabel={$\numChainEvents$},
grid=both,
clip=false,
axis lines*=left,
title={Insert, $\numThreads=10$},
ymin={0},
xtick distance=0.25e6,
ytick distance=0.25e-4,
]

\addplot[mark=*, mark size=2pt,color=black] table[x=x, y=cssts, col sep=comma]{scalability_sts_insert_k=10_r=10000_i=20n.csv};
\addplot[mark=diamond*, mark size=3pt,color=olive] table[x=x, y=sts, col sep=comma]{scalability_sts_insert_k=10_r=10000_i=20n.csv};\label{plots:curve_sts}

\nextgroupplot[
inner sep=2pt,
width=\width,
height=\height,
ylabel={Time (s)},
ylabel near ticks,
xlabel={$\numChainEvents$},
grid=both,
clip=false,
axis lines*=left,
title={Insert, $\numThreads=20$},
ymin={0},
xtick distance=0.25e6,
ytick distance=1e-4,
]

\addplot[mark=*, mark size=2pt,color=black] table[x=x, y=cssts, col sep=comma]{scalability_sts_insert_k=20_r=10000_i=20n.csv};
\addplot[mark=diamond*, mark size=3pt,color=olive] table[x=x, y=sts, col sep=comma]{scalability_sts_insert_k=20_r=10000_i=20n.csv};

\nextgroupplot[
inner sep=2pt,
width=\width,
height=\height,
ylabel={Time (s)},
ylabel near ticks,
xlabel={$\numChainEvents$},
grid=both,
clip=false,
axis lines*=left,
title={Query, $\numThreads=10$},
ymin={0},
xtick distance=0.25e6,
ytick distance=1e-6,
]

\addplot[mark=*, mark size=2pt,color=black] table[x=x, y=cssts, col sep=comma]{scalability_sts_query_k=10_r=10000_i=20n.csv};
\addplot[mark=diamond*, mark size=3pt,color=olive] table[x=x, y=sts, col sep=comma]{scalability_sts_query_k=10_r=10000_i=20n.csv};

\nextgroupplot[
inner sep=2pt,
width=\width,
height=\height,
ylabel={Time (s)},
ylabel near ticks,
xlabel={$\numChainEvents$},
grid=both,
clip=false,
axis lines*=left,
title={Query, $\numThreads=20$},
ymin={0},
xtick distance=0.25e6,
ytick distance=2e-6,
legend to name=named
]

\addplot[mark=*, mark size=2pt,color=black] table[x=x, y=cssts, col sep=comma]{scalability_sts_query_k=20_r=10000_i=20n.csv};
\addlegendimage{mark=square*, mark size=2pt,color=teal}
\addplot[mark=diamond*, mark size=3pt,color=olive] table[x=x, y=sts, col sep=comma]{scalability_sts_query_k=20_r=10000_i=20n.csv};
\coordinate (bot) at (rel axis cs:1,0);% coordinate at bottom of the last plot

\legend{$\ds$, $\vc$, $\naiveds$}
\end{groupplot}
\path (top)--(bot) coordinate[midway] (group center);
\node[inner sep=0pt, below=0.15em] at(group center |- current bounding box.south) {\pgfplotslegendfromname{named}};
\end{tikzpicture}
}%
\caption{
Scalability experiments on $\ds$, $\naiveds$ and $\vc$.
Each partial order contains $\numThreads$ chains and $\numChainEvents=\numEvents/\numThreads$ events per chain.
}
\label{fig:scalability}
\end{figure*}

%% file: sections/related_work-main.tex
\section{Related Work}
The efficient maintenance of partial orders is a key component in concurrent program analysis.
%and has been studied in various contexts.
Vector Clocks~\cite{Mattern-1989} were originally proposed as an efficient mechanism for causality tracking in distributed systems~\cite{Lamport-1978}, and have been used extensively in program analysis.
Their efficient implementation has been approached in various ways, e.g., using varying-size arrays~\cite{Christiaens-2001}, AVL trees~\cite{Li-2019}, Chain Clocks~\cite{Agarwal-2005}, and Tree Clocks~\cite{Mathur-2022}.
These variations offer performance improvements over vanilla Vector Clocks, but only when the partial order is built incrementally and in a streaming fashion.

For non-streaming updates, partial orders are normally represented as plain graphs~\cite{Norris-2013,Biswas-2014,Abdulla-2019,Roemer-2018,Cirisci-2020,Zennou-2020} and Vector Clocks~\cite{Luo-2021,Gao-2023}.
The closest data structure to $\ds$ has been $\naiveds$, developed in~\cite{Pavlogiannis-2019} for  data-race detection, and subsequently used in other dynamic analyses~\cite{Cai-2021,Yu-2021,Shi-2024}.
The development of $\ds$ lies on a few technical novelties.
For example, our minima indexing and sparse tree representation (\cref{subsec:sparse_segment_tree}) are novel, and lead to tighter complexity bounds.
Moreover, in order to handle the fully dynamic setting, $\ds$ do not store transitive reachability.
This requires overcoming the challenge of discovering transitive reachability during queries in an efficient manner (\cref{subsec:ds-cssts}).

%% file: sections/conclusion.tex
\section{Conclusion}\label{sec:conclusion}

We have introduced $\ds$, a data structure for the fully-dynamic maintenance of partial orders with small width, supporting a plethora of dynamic analyses for concurrency.
%$\ds$ allow inserting and deleting new orderings, as well as querying for existing orderings, each in logarithmic time, and is even more efficient when the partial order adheres to certain sparsity conditions, which are common in practice.
Our experimental results indicate that $\ds$ are an efficient, drop-in replacement of existing approaches to maintaining partial orders in various application domains of recent literature.
Our aim is for $\ds$ to become a standard reference for driving future scalable dynamic analyses.

%% file: artifact_appendix.tex
%!TEX root = main.tex

% LaTeX template for Artifact Evaluation V20201122
%
% Prepared by Grigori Fursin with contributions from Bruce Childers,
%   Michael Heroux, Michela Taufer and other colleagues.
%
% See examples of this Artifact Appendix in
%  * SC'17 paper: https://dl.acm.org/citation.cfm?id=3126948
%  * CGO'17 paper: https://www.cl.cam.ac.uk/~sa614/papers/Software-Prefetching-CGO2017.pdf
%  * ACM ReQuEST-ASPLOS'18 paper: https://dl.acm.org/citation.cfm?doid=3229762.3229763
%
% (C)opyright 2014-2022
%
% CC BY 4.0 license
%

%\documentclass{sigplanconf}

%\usepackage{hyperref}

%\begin{document}
	
%	\special{papersize=8.5in,11in}
	
	%%%%%%%%%%%%%%%%%%%%%%%%%%%%%%%%%%%%%%%%%%%%%%%%%%%%
	% When adding this appendix to your paper, 
	% please remove above part
	%%%%%%%%%%%%%%%%%%%%%%%%%%%%%%%%%%%%%%%%%%%%%%%%%%%%
	
	%\appendix
	\section{Artifact Appendix}\label{sec:artifact_appendix}

	%%%%%%%%%%%%%%%%%%%%%%%%%%%%%%%%%%%%%%%%%%%%%%%%%%%%%%%%%%%%%%%%%%%%%
	\subsection{Abstract}
	
	This artifact contains all the source codes and experimental data required to replicate our evaluation in~\cref{sec:experiments}. 
	In particular, it contains the following content: (i) source code of the data structures $\ds$, $\naiveds$, $\vc$, and $\graphds$ (ii) source code of the analyses used in generating \crefrange{tbl:m2}{tbl:rc} and the corresponding experimental data.
	We additionally provide Python scripts that automate the process of replicating our results.

 Besides the accompanying artifact, $\ds$ are available in a repository~\cite{CSSTsRepot} as a stand alone data structure for dynamic graph reachability that can be used in other settings.
	
	\subsection{Artifact check-list (meta-information)}
		
	{\small
		\begin{itemize}
			\item {\bf Algorithm: } Collective Sparse Segment Trees (CSSTs).
			\item {\bf Run-time environment: }  Docker. 
			\item {\bf Metrics: } Execution time.
			\item {\bf Output: } CSV files.
			\item {\bf How much disk space required (approximately)?: } 20 GB.
			\item {\bf How much time is needed to prepare workflow (approximately)?: }  We provide all the scripts that automate our workflow.
			\item {\bf How much time is needed to complete experiments (approximately)?: } 80 hours. We also provide the users with the option to run the experiments on a smaller set of benchmarks or use a shorter timeout.
			\item {\bf Publicly available?: } Yes \cite{artifact}.
			\item {\bf Code licenses (if publicly available)?: } MIT License.
			\item {\bf Archived (provide DOI)?: } \href{https://doi.org/10.5281/zenodo.10798906}{10.5281/zenodo.10798906}
		\end{itemize}
	}
	
	%%%%%%%%%%%%%%%%%%%%%%%%%%%%%%%%%%%%%%%%%%%%%%%%%%%%%%%%%%%%%%%%%%%%%
	\subsection{Description}
	
	\subsubsection{How to access}
	Obtain the artifact from~\cite{artifact}.
	
	\subsubsection{Hardware dependencies}
	Replicating the results of large benchmarks requires up to 60 GB RAM.
	Moreover, we remark that users may encounter problems in running the artifact on Apple M1 silicon due to certain incompatibility issues in running Docker on these machines.
	There are otherwise no special hardware requirements.
	
	\subsubsection{Software dependencies}
	Docker.
	
	\subsubsection{Data sets}
	All data sets are provided with the artifact.

	%%%%%%%%%%%%%%%%%%%%%%%%%%%%%%%%%%%%%%%%%%%%%%%%%%%%%%%%%%%%%%%%%%%%%
	\subsection{Installation}
	
	\begin{itemize}
		\item Install Docker (\href{https://www.docker.com}{https://www.docker.com}).
		\item Obtain the artifact's Docker image from~\cite{artifact}.
		\item { Import the image: 
			\begin{verbatim}
				> docker import asplos24-139.tar asplos24-139
			\end{verbatim}
		}
		\item {
			Start a container:
			\begin{verbatim}
				> docker run -it asplos24-139 bash
			\end{verbatim}
		}
	\end{itemize}
	
	%%%%%%%%%%%%%%%%%%%%%%%%%%%%%%%%%%%%%%%%%%%%%%%%%%%%%%%%%%%%%%%%%%%%%
	\subsection{Experiment workflow}
	
	\cref{fig:artifact-structure} displays the directory structure of our artifact. 
	The directory \texttt{src} contains the source code of the data structures $\ds$, $\naiveds$, and $\vc$. 
	The directory \texttt{experiment} is the primary folder for experimental evaluation.
	The subdirectories \texttt{analyses} and \texttt{data} contain the source code of the analyses and their corresponding experimental data, respectively.
	The directories \texttt{output} and \texttt{result} serve as destinations for the generated outputs and are initially empty.
	The file \texttt{experiment.py} defines the experimental workflow. 
	Users do not need to interact directly with this program.
	The files \texttt{run.py} and \texttt{compile\_results.py} are helper scripts that interface with \texttt{experiment.py} to execute the experiments and generate results in a tabular format.

	\begin{figure}[h]
		\centering
		\noindent
		\begin{verbatim}
			root/
			|-- src/
			|-- experiment/
			    |-- analyses/
			    |-- data/
			    |-- output/
			    |-- result/
			    |-- experiment.py
			    |-- run.py
			    |-- compile_results.py
		\end{verbatim}
		\caption{Directory structure of the artifact}
		\label{fig:artifact-structure}
	\end{figure}
	
	%%%%%%%%%%%%%%%%%%%%%%%%%%%%%%%%%%%%%%%%%%%%%%%%%%%%%%%%%%%%%%%%%%%%%
	\subsection{Evaluation and expected results}
	Running the script \texttt{run.py} will execute all the experiments.
	\begin{verbatim}
		> python3 run.py
	\end{verbatim}
	The generated raw outputs can be located under the folder \texttt{output}.
	The file \texttt{compile\_results.py} can be used to generate outputs in a tabular format.
	\begin{verbatim}
	> python3 compile_results.py
	\end{verbatim}
	This will generate a \texttt{CSV} file for each analyses under the folder \texttt{results}.
	The contents of these files can be displayed in the console as follows.
	\begin{verbatim}
		> csvtool readable /root/experiments/results/<NAME>.csv
	\end{verbatim}
	The main goal of this evaluation is to measure the performance
	benefits of $\ds$ over $\naiveds$, $\vc$ and $\graphds$ in various dynamic concurrency analyses. 
	We expect that for each analyses the overall speedups would remain similar to the results reported in \crefrange{tbl:m2}{tbl:rc}.
	
	%%%%%%%%%%%%%%%%%%%%%%%%%%%%%%%%%%%%%%%%%%%%%%%%%%%%%%%%%%%%%%%%%%%%%
	\subsection{Experiment customization}
	Users may customize experiments by passing arguments to the script \texttt{run.py}. 
	For instance, executing the following command will run the experiments on a smaller subset of benchmarks:
	\begin{verbatim}
		> python3 run.py -s 1
	\end{verbatim}
	Whereas the following command sets the timeout for each individual experiment to $30$ minutes.
	\begin{verbatim}
		> python3 run.py -t 30m
	\end{verbatim}
	For a comprehensive list of supported options, please refer to the documentation accompanying the script.
	\begin{verbatim}
		> python3 run.py --help
	\end{verbatim}
%\end{document}

%% file: appendix/0-main.tex
\input{appendix/proofs}
\input{appendix/app_experiments}

%% file: appendix/proofs.tex
\section{Proofs}\label{sec:app_proofs}

In this section we provide proofs of the formal statements of the main paper.

\lemsparsetreeheight*
\begin{proof}
Observe that every intermediate node $\node_{i,j}$ of a $\sparsest$ $T$  has its field $\node_{i,j}.\pos$ point to a non-empty entry of $A[i,j]$.
Due to \cref{eq:pos}, this is a different entry from those pointed by the $\pos$ fields of the ancestors of $\node_{i,j}$.
Moreover, any node $\node_{i',j'}$ which is neither an ancestor nor a descendant of $\node_{i,j}$ has a non-overlapping range with $\node_{i,j}$, i.e., $[i:j]\cap[i':j']=\emptyset$.
It follows that $\node_{i,j}.\pos$ indexes a unique entry of $A$.
In turn this implies that the height of $T$ is bounded by the number $d$ of non-empty elements of $A$.

In addition, as in standard Segment Trees, for every two nodes $\node_{i,j}$ and $\node_{i', j'}$ such that $\node_{i,j}$ is the parent of $\node_{i',j'}$, we have that 
$(j'-i')\leq \lceil (j-i)/2 \rceil$, i.e., the range halves in each step.
Thus the height $T$ is also bounded by $\log \numEvents$, as desired.
\end{proof}

\lemdssparsity*
\begin{proof}
Since we have $\dsarray{t_1}{t_2}[j_1]\neq \infty$ iff there exists some $j_2\in [\numEvents]$ with $\graphnode{t_1}{j_1}\to \graphnode{t_2}{j_2}$,
it follows that the density of $\dsarray{t_1}{t_2}$ is bounded by the cross-chain density $\CrossChainDensity$ of $G$.
\end{proof}

\lemdsinvariant*
\begin{proof}
The statement follows from the invariant that $\dsarray{t_1}{t_2}[j_1]$ is equal to the root of the min-heap $\edgeheap_{t_1}^{t_2}[j_1]$, which, at all times, contains all $j_2\in[\numEvents]$ such that $\graphnode{t_1}{j_1}\to \graphnode{t_2}{j_2}$.
\end{proof}

\lemdscomplexitysuc*
\begin{proof}
The statement follows by induction on $i$, in a style similar to correctness proof of the Bellman-Ford algorithm.
In particular, for $i=0$, every crossing path consists of two nodes $\pi :\graphnode{t_1}{j_1},\graphnode{t_2}{j_2}$, and indeed, $\closure_{t_1}^{t_1}[j_1]=j_1$, by initialization loop in \cref{line:ds-full_foreach}.

Now assume that the statement holds for some $i$, and we argue that it holds for $i+1$.
Consider any crossing path $\pi :\graphnode{t_1}{j_1},\dots, \graphnode{t_2}{j_2}$ of length $i+1$, and let $\pi'$ be its prefix of length $i$, ending in some node $\graphnode{t_3}{j_3}$.
By the induction hypothesis, we have that after the $i$-th iteration of the do-while loop, we have $\closure_{t_1}^{t_3}[j_1]\leq j_3$.
Moreover, since the last pair of $\pi$ is $\dots \graphnode{t_3}{j_3},\graphnode{t_2}{j_2}$, there exist $i_1, i_2\in[\numThreads]$ with $i_1\geq j_3$ and $i_2\leq j_2$ such that $\graphnode{t_3}{i_1}\to \graphnode{t_2}{i_2}$.
Then the suffix minima query in \cref{line:ds-full-v} will obtain $v\leq i_2$, and thus the invariant is restored in \cref{line:ds-full-update_closure}.
\end{proof}

\lemdsincsoundness*
\begin{proof}
Consider any update $\upd(\dsarray{t_{1}'}{t_{2}'}, j_{1}', j_{2}')$ happening in \cref{line:ds-inc-interface-ins-8} when the algorithm processes a call $\insedge(\graphnode{t_1}{j_1}, \graphnode{t_2}{j_2})$, and we will argue that $\graphnode{t'_1}{j'_1}\reach \graphnode{t'_2}{j'_2}$.
Then $\graphnode{t'_1}{j'_1}\reach \graphnode{t_1}{j_1}$ and $\graphnode{t_2}{j_2}\reach \graphnode{t'_2}{j'_2}$, thus after the end $\graphnode{t_1}{j_1}\to\graphnode{t_2}{j_2}$ is inserted in $G$, we indeed have $\graphnode{t'_1}{j'_1}\reach \graphnode{t'_2}{j'_2}$, as desired.
\end{proof}

\lemdsinccompleteness*
\begin{proof}
Consider any two nodes $\graphnode{t'_1}{j'_1}\reach\graphnode{t'_2}{j'_2}$ after the algorithm processes a call $\insedge(\graphnode{t_1}{j_1}, \graphnode{t_2}{j_2})$, and we will argue that there exists some $i'_2\in [\numEvents]$ such that $i'_2\leq j'_2$ and $\dsarray{t'_1}{t'_2}[j_1]=i'_2$.
The statement holds by induction if there is a path $\graphnode{t'_1}{j'_1}\reach\graphnode{t'_2}{j'_2}$ not containing the edge $\graphnode{t_1}{j_1}\to\graphnode{t_2}{j_2}$.
Otherwise, we have 
(i)~a path $\graphnode{t_2}{j_2}\reach \graphnode{t'_2}{j'_2}$, and
(ii)~a path $\graphnode{t'_1}{j'_1}\reach \graphnode{t'_1}{j'_1}$.
These two paths are merged appropriately in \cref{line:ds-inc-interface-ins-8}, leading to $\dsarray{t'_1}{t'_2}[j_1]=i'_2$, as desired.
\end{proof}

\lemdsincsparsity*
\begin{proof}
The statement follows from the following observation.
Whenever \cref{algo:ds-inc-interface} inserts an edge $\upd(\dsarray{t_{1}'}{t_{2}'}, j_{1}', j_{2}')$,
we have that either
(i) $t'_1=t_1$, in which case $j'_1=j_1$, or
(ii) $j_1=\pred(\graphnode{t_1}{ j_1}, t_{1}')$.
In the first case we are inserting an outgoing edge to $\graphnode{t'_1}{j'_1}$, while in the second case, $\graphnode{t'_1}{j'_1}$ already has an edge to some other chain.
In both cases, $\graphnode{t'_1}{j'_1}$ counts towards the cross-chain density $\CrossChainDensity$ of $G$.
Hence, the number of non-$\infty$ entries of each suffix-minima array is bounded by $\CrossChainDensity$, as desired.
\end{proof}

%% file: appendix/app_experiments.tex
\section{Experiment Details}\label{sec:app_experiments}

\input{appendix/experiments_full_tables}

%% file: appendix/experiments_full_tables.tex
%!TEX root = ../main.tex

\subsection{Complete Experimental Results}\label{subsec:app_full_tables}

For the sake of completeness, here we present the full versions of \cref{tbl:m2,tbl:dl,tbl:cc,tbl:c11tester}, including benchmark entries on which all data structures yielded a runtime of $<1$s.

\newpage

\input{material/tables/m2-result-full}

\input{material/tables/deadlock-result-full}

\input{material/tables/consistency-checking-result-full}

\input{material/tables/c11tester-result-full}

%% file: material/tables/m2-result-full.tex
%!TEX root = ../../main.tex

\begin{table}[h!]
    \caption{
    %Race prediction results. Columns 1-2 denote the number of threads and events, respectively.
    %Columns 3-5 show the times taken in seconds. Timeout is set to 5h.
    Race prediction results. The timeout is 5h.
    }
    %\vspace{-0.15cm}
    \setlength\tabcolsep{3pt}
    \small
    \centering
    \scalebox{1}{
    \begin{tabular}{|r|c|c|c|c|c|c|}
\hline
\textbf{benchmark} & $\mathit{T}$ & $\mathit{N}$ & $\AvgDensity$ & $\vc$ (s) & $\naiveds^\dagger$ (s) & $\ds$ (s)\\
\hline\hline
array & 6 & 63 & 0.25 & 0.1 & 0.1 & 0.1\\
critical & 7 & 73 & 0.26 & 0.1 & 0.1 & 0.1\\
account & 7 & 148 & 0.23 & 0.1 & 0.1 & 0.1\\
airlinetickets & 14 & 168 & 0.22 & 0.1 & 0.1 & 0.1\\
pingpong & 21 & 193 & 0.17 & 0.1 & 0.1 & 0.1\\
twostage & 15 & 223 & 0.22 & 0.1 & 0.1 & 0.1\\
wronglock & 25 & 296 & 0.3 & 0.8 & 0.9 & 0.7\\
boundedbuffer & 7 & 346 & 0.2 & 0.1 & 0.1 & 0.1\\
prodcons & 11 & 680 & 0.12 & 0.1 & 0.1 & 0.1\\
clean & 12 & 1.3K & 0.28 & 2.4 & 2.6 & 1.6\\
mergesort & 8 & 3.0K & 0.19 & 0.3 & 0.3 & 0.3\\
bubblesort & 29 & 4.7K & 0.15 & 69.4 & 56.2 & 27.5\\
lang & 10 & 6.3K & 0.22 & 93.3 & 39.2 & 27.3\\
readerswriters & 8 & 11.3K & 0.32 & 54.4 & 23.5 & 18.5\\
raytracer & 6 & 15.8K & 0.17 & 10.5 & 10.2 & 9.3\\
bufwriter & 9 & 22.3K & 0.2 & 130 & 24.6 & 12.9\\
ftpserver & 14 & 49.6K & 0.06 & 41.2 & 14.1 & 9.0\\
moldyn & 6 & 200.3K & 0.12 & T.O. & T.O. & 12636\\
linkedlist & 15 & 1.0M & 0.06 & T.O. & T.O. & 8893\\
derby & 7 & 1.4M & 0.04 & 1436 & 215 & 197\\
jigsaw & 15 & 3.1M & 0.06 & 154 & 45.6 & 32.0\\
sunflow & 17 & 11.7M & 0.01 & T.O. & 780 & 505\\
xalan & 9 & 122.5M & 0.01 & T.O. & O.O.M. & 979\\
batik & 8 & 157.9M & 0.01 & O.O.M. & O.O.M. & 1956\\
\hline\hline\textbf{Total} & \textbf{-} & \textbf{297.9M} & \textbf{-} & \textbf{>91994} & \textbf{>73213} & \textbf{25305}\\
                    \hline
                \end{tabular}
            }
        \label{tbl:m2-full}
        \end{table}

%% file: material/tables/deadlock-result-full.tex
%!TEX root = ../../main.tex

\begin{table}[h!]
    \caption{
    %Consistency checking results.  Columns 1-2 denote the number of threads and events, respectively.
    %Columns 3-5 show the times taken in seconds. Timeout is set to 1h.
    Deadlock results. %The timeout is 1h.
    }
    %\vspace{-0.15cm}
    \setlength\tabcolsep{3pt}
    \small
    \centering
    \scalebox{1}{
    \begin{tabular}{|r|c|c|c|c|c|c|}
\hline
\textbf{benchmark} & $\mathit{T}$ & $\mathit{N}$ & $\AvgDensity$ & $\vc$ (s) & $\naiveds^\dagger$ (s) & $\ds$ (s)\\
\hline\hline
Deadlock & 3 & 37 & 0.32 & 0.1 & 0.1 & 0.1\\
NotADeadlock & 3 & 54 & 0.13 & 0.1 & 0.1 & 0.1\\
Picklock & 3 & 60 & 0.16 & 0.1 & 0.1 & 0.1\\
Bensalem & 4 & 63 & 0.35 & 0.1 & 0.1 & 0.1\\
Transfer & 3 & 66 & 0.11 & 0.1 & 0.1 & 0.1\\
Test-Dimminux & 3 & 67 & 0.15 & 0.1 & 0.1 & 0.1\\
StringBuffer & 3 & 72 & 0.24 & 0.1 & 0.1 & 0.1\\
Test-Calfuzzer & 5 & 156 & 0.04 & 0.1 & 0.1 & 0.1\\
DiningPhil & 6 & 272 & 0.04 & 0.1 & 0.1 & 0.1\\
HashTable & 3 & 316 & 0.29 & 0.2 & 0.2 & 0.2\\
Log4j2 & 4 & 1.5K & 0.04 & 0.2 & 0.2 & 0.2\\
Dbcp1 & 3 & 2.2K & 0.09 & 0.2 & 0.2 & 0.2\\
Dbcp2 & 3 & 2.5K & 0.03 & 0.2 & 0.2 & 0.2\\
Derby2 & 3 & 2.5K & 0.07 & 0.2 & 0.2 & 0.2\\
RayTracer & 5 & 30.6K & 0.05 & 0.5 & 0.5 & 0.5\\
jigsaw & 21 & 143.0K & 0.03 & 356 & 14.8 & 7.8\\
elevator & 5 & 245.9K & 0.02 & 3.7 & 2.8 & 2.6\\
hedc & 7 & 409.8K & 0.04 & 23.2 & 22.0 & 23.2\\
JDBCMySQL-1 & 3 & 442.0K & 0.01 & 3.0 & 3.0 & 2.8\\
JDBCMySQL-2 & 3 & 442.1K & 0.01 & 3.1 & 2.9 & 3.0\\
JDBCMySQL-3 & 3 & 442.8K & 0.01 & 2.8 & 3.1 & 3.2\\
JDBCMySQL-4 & 3 & 442.9K & 0.01 & 3.7 & 4.1 & 3.5\\
cache4j & 2 & 775.5K & 0.01 & 5.5 & 6.6 & 5.6\\
Swing & 8 & 3.8M & 0.01 & 14.8 & 13.4 & 13.1\\
sunflow & 15 & 21.5M & 0.01 & 369 & 111 & 110\\
eclipse & 15 & 64.2M & 0.01 & 535 & 711 & 282\\
\hline\hline\textbf{Total} & \textbf{-} & \textbf{92.8M} & \textbf{-} & \textbf{1322} & \textbf{897} & \textbf{459}\\
        \hline
        \end{tabular}
        }
    \label{tbl:dl-full}
    \end{table}

%% file: material/tables/consistency-checking-result-full.tex
%!TEX root = ../../main.tex

\begin{table}[h!]
    \caption{
    %Consistency checking results.  Columns 1-2 denote the number of threads and events, respectively.
    %Columns 3-5 show the times taken in seconds. Timeout is set to 1h.
    Consistency checking results. The timeout is 1h.
    }
    %\vspace{-0.15cm}
    \setlength\tabcolsep{3pt}
    \small
    \centering
    \scalebox{1}{
    \begin{tabular}{|r|c|c|c|c|c|c|}
\hline
\textbf{benchmark} & $\mathit{T}$ & $\mathit{N}$ & $\AvgDensity$ & $\vc$ (s) & $\naiveds$ (s) & $\ds$ (s)\\
\hline\hline
dekker & 3 & 16.3K & 0.41 & 6.9 & 0.3 & 0.25\\
sigma & 10 & 18.0K & 0.13 & 1.0 & 0.4 & 0.3\\
peterson & 3 & 19.0K & 0.37 & 9.4 & 0.4 & 0.45\\
lamport & 3 & 28.1K & 0.39 & 18.3 & 0.6 & 0.6\\
dq & 4 & 31.5K & 0.24 & 11.5 & 0.5 & 0.44\\
chase-lev & 5 & 32.2K & 0.24 & 23.3 & 0.6 & 0.51\\
control-flow & 27 & 39.0K & 0.01 & 0.9 & 0.7 & 0.64\\
szymanski & 3 & 77.9K & 0.33 & 17.4 & 0.9 & 0.8\\
buf-ring & 9 & 115.3K & 0.39 & 322 & 7.3 & 6.27\\
mcs-lock & 11 & 196.4K & 0.17 & 298 & 28.3 & 20.71\\
spsc & 3 & 243.6K & 0.53 & 2548 & 2.1 & 1.74\\
linuxrwlocks & 6 & 276.4K & 0.28 & T.O. & 12.6 & 11.18\\
fib-bench & 3 & 300.0K & 0.4 & T.O. & 4.7 & 4.21\\
seqlock & 17 & 318.1K & 0.15 & 1227 & 46.1 & 28.16\\
spinlock & 11 & 482.5K & 0.39 & T.O. & 103 & 75.11\\
ttaslock & 11 & 491.1K & 0.41 & T.O. & 111 & 81.19\\
exp-bug & 4 & 498.5K & 0.38 & 2591 & 3.6 & 2.76\\
mutex & 11 & 519.7K & 0.56 & T.O. & 122 & 86.09\\
ticketlock & 6 & 569.6K & 0.4 & T.O. & 20.8 & 17.17\\
gcd & 3 & 750.1K & 0.28 & T.O. & 6.0 & 4.52\\
indexer & 17 & 800.0K & 0.51 & 7.8 & 5.8 & 3.23\\
twalock & 11 & 900.0K & 0.43 & T.O. & 174 & 118.39\\
treiber & 6 & 1.0M & 0.22 & T.O. & 20.7 & 16.73\\
mpmc & 10 & 2.0M & 0.17 & T.O. & 41.6 & 22.59\\
barrier & 5 & 2.8M & 0.6 & T.O. & 139 & 112.61\\
\hline\hline\textbf{Total} & \textbf{-} & \textbf{12.5M} & \textbf{-} & \textbf{>46682} & \textbf{853} & \textbf{617}\\
        \hline
        \end{tabular}
        }
    \label{tbl:cc-full}
    \end{table}

%% file: material/tables/c11tester-result-full.tex
%!TEX root = ../../main.tex

\begin{table}[h!]
    \caption{
    Race detection on C11.
    %$\celeventester$ results.  Columns 1-2 denote the number of threads and events, respectively.
    %Columns 3-5 show the times taken in seconds.
    %The benchmarks in which propagation is observed are marked with (*).
    }
    %\vspace{-0.15cm}
    \setlength\tabcolsep{3pt}
    \small
    \centering
    \scalebox{1}{
    \begin{tabular}{|r|c|c|c|c|c|c|}
\hline
\textbf{benchmark} & $\mathit{T}$ & $\mathit{N}$ & $\AvgDensity$ & $\vc^\dagger$ (s) & $\naiveds$ (s) & $\ds$ (s)\\
\hline\hline
buf-ring & 10 & 48.7K & 0.25 & 0.3 & 0.3 & 0.3\\
dq & 5 & 72.8K & 0.48 & 2.5 & 2.8 & 3.1\\
chase-lev & 6 & 91.0K & 0.35 & 0.2 & 0.2 & 0.2\\
mabain & 7 & 101.6K & 0.33 & 0.6 & 1.2 & 1.1\\
peterson & 4 & 139.6K & 0.67 & 0.1 & 0.2 & 0.1\\
lamport & 4 & 201.3K & 0.42 & 0.2 & 0.2 & 0.2\\
dekker & 4 & 219.2K & 0.42 & 0.3 & 0.3 & 0.3\\
control-flow & 28 & 260.0K & 0.01 & 0.5 & 0.2 & 0.2\\
sigma & 11 & 360.0K & 1 & 0.9 & 0.4 & 0.5\\
seqlock & 18 & 693.9K & 0.2 & 4.5 & 6.7 & 6.3\\
iris-1 & 13 & 1.1M & 0.2 & 54.8 & 66.5 & 60.3\\
qu & 11 & 1.3M & 0.43 & 5.9 & 7.5 & 7.2\\
indexer & 18 & 1.6M & 1 & 1.0 & 1.1 & 1.1\\
szymanski & 4 & 1.9M & 0.52 & 0.8 & 0.9 & 0.9\\
exp-bug & 5 & 2.5M & 0.25 & 2.6 & 2.5 & 2.5\\
twalock & 12 & 4.5M & 0.34 & 29.6 & 53.3 & 49.0\\
gcd & 4 & 4.5M & 1 & 2.6 & 2.8 & 2.7\\
spinlock & 12 & 4.8M & 0.2 & 23.0 & 37.9 & 34.0\\
ttaslock & 12 & 4.9M & 0.25 & 26.7 & 48.0 & 42.7\\
silo & 5 & 5.6M & 0.01 & 5.9 & 6.2 & 6.1\\
fib-bench & 4 & 6.0M & 1 & 6.5 & 6.3 & 6.5\\
linuxrwlocks & 7 & 6.9M & 0.39 & 30.8 & 40.6 & 37.7\\
barrier & 6 & 8.3M & 0.51 & 19.6 & 24.5 & 23.7\\
mpmc & 11 & 9.2M & 0.24 & 37.6 & 44.9 & 42.6\\
spsc & 4 & 9.7M & 0.55 & 12.2 & 12.2 & 12.1\\
mcs-lock & 12 & 9.9M & 0.37 & 55.2 & 86.8 & 80.1\\
treiber & 7 & 10.1M & 0.11 & 31.3 & 36.6 & 35.7\\
iris-2 & 4 & 11.5M & 0.2 & 15.5 & 14.7 & 14.7\\
gdax & 8 & 13.5M & 0.97 & 8.5 & 7.6 & 7.6\\
ticketlock & 7 & 14.2M & 0.48 & 45.1 & 70.4 & 64.2\\
mutex & 12 & 15.6M & 0.39 & 48.5 & 68.5 & 63.7\\
readerswriters & 13 & 10.1M & 0.33 & 16.3 & 7.6 & 7.5\\
atomicblocks & 33 & 15.5M & 0.5 & 9.7 & 1.7 & 1.6\\
\hline\hline\textbf{Total} & \textbf{-} & \textbf{175.6M} & \textbf{-} & \textbf{500} & \textbf{662} & \textbf{617}\\
            \hline
            \end{tabular}
            }
        \label{tbl:c11tester-full}
        \end{table}